\documentclass[prd,reprint]{revtex4-1}

\usepackage{amsmath}
\usepackage{amsfonts}
\usepackage{amssymb}
\usepackage{graphicx}
\usepackage{epstopdf}
\usepackage[usenames,dvipsnames]{color}
\usepackage{soul}

\def\se#1{\begin{equation}#1\end{equation}}
\def\sen#1{\begin{equation*}#1\end{equation*}}

\def\R{\mathbb{R}}
\def\RP2{\mathbb{R}\textrm{P}^2}
\def\braket#1{\left\langle#1\right\rangle}

\newcommand{\hll}{}
\newcommand{\stt}[1]{}
\newcommand{\hlref}{}

\begin{document}

\title{Statistical mechanics of graph models\\
and their implications for emergent manifolds}

\date{\today}

\author{Si~Chen}
\author{Steven~S.~Plotkin}
\affiliation{Department of Physics and Astronomy, University of British Columbia, Vancouver, Canada V6T 1Z1}

\begin{abstract}
Inspired by ``quantum graphity'' models for spacetime, a statistical
model of graphs is proposed to explore possible realizations of
emergent manifolds. Graphs with given numbers of vertices and edges
are considered, governed by a very general Hamiltonian that merely favors
graphs with near-constant valency and local rotational symmetry. The
ratio of vertices to edges controls the dimensionality of the emergent
manifold. The model is simulated numerically in the canonical ensemble
for a given vertex to edge ratio, where it is found that the low-energy states are almost triangulations of two-dimensional
manifolds. The resulting manifold shows topological ``handles''
and surface intersections in a higher embedding space, as well as
non-trivial fractal dimension \hll{consistent with previous spectral
analysis, and nonlocal links consistent with models of disordered
locality}. The
transition \hll{to an emergent manifold} is first order,
\hll{and thus dependent on microscopic structure.}
\stt{underlying the difficulty of graph
models in describing criticality that is independent of the details of
the underlying graph.}
\hll{Issues involved in interpreting nearly-fixed valency graphs 
as Feynman diagrams dual to a triangulated manifold as in matrix models are discussed.}
Another interesting phenomenon is that the
entropy of the graphs are super-extensive, a fact known since Erd\H{o}s,
which results in a transition temperature of zero in the limit of
infinite system size: infinite manifolds are always disordered. Aside
from a finite universe or diverging coupling constraints
as possible solutions to this problem, long-range interactions
between vertex defects also resolve the problem and restore a nonzero
transition temperature, in a manner similar to that in low-dimensional
condensed-matter systems.
\end{abstract}

\maketitle

\section{Introduction}

Since the first systematic studies of random graph models by Erd\H{o}s
and R\'{e}nyi \cite{Erdos:1959}, the relation between graph theory
models and physics models, in particular statistical physics models,
has  attracted much interest.
Concepts and tools in graph theory have been applied to problems in
physics, computer science, and biology to produce remarkable results.
For example, Feynman diagrams that are planar have special roles in the large
$N$ QCD model~\cite{'tHooft:1973jz};
in causal dynamical triangulation, four-dimensional triangulated manifolds with fixed edge lengths, which can be viewed as a class of graphs, are used to construct
spacetime on the Planck scale to regularize the quantum gravitational
path integral~\cite{Ambjorn:2004qm,Ambjorn:2005qt};
statistical mechanical models of network growth can explain the
connectivity of systems such as the Internet~\cite{Albert:2002zz}; structures of amorphous solids
can be quantified using graph theory properties~\cite{Franzblau:1991};
intracellular signalling networks can exhibit emergent behavior stored
within biochemical reactions, including integration
of signals across multiple time scales and self-sustaining feedback
loops~\cite{Bhalla:1999}; neural networks can collectively and robustly produce
content-addressable memories from partial cues~\cite{Hopfield:1982},
indicating capacity for generalization, familiarity recognition, and
categorization. Added to these discoveries, a new collection of graph models
has been proposed as candidates for emergent spacetime, as described
below.

A manifold can be approximated by a triangulation, which in turn can be viewed as a graph filled with simplices. From this observation, one can consider
how a graph may give rise to a manifold; i.e. from a family of graphs, following some constraints and
obeying some set of rules for dynamical processes, is it possible that a manifold-like
structure can emerge? To be more precise, consider the possibility
that a graph $G$ gives rise to a smooth manifold $M$. A vertex in $G$
corresponds to a point in $M$; when a pair of vertices in $G$ are
connected by an edge, the corresponding pair of points in $M$ have a
certain distance $\epsilon$. When the length scale under consideration
is much larger than $\epsilon$, $G$ 
resembles the smooth manifold
$M$. In such cases, one can say that the manifold $M$, including its
dimensionality, topology, and metric, emerges from the graph $G$ in
the continuous limit.

From this general idea, in references
\cite{Konopka:2006hu,Konopka:2008hp}, a graph model was constructed
from a given graph Hamiltonian,
where it was proposed that the low-energy phase of the model may be interpreted as an emergent
spacetime. In addition, it was found that when the edges of the graph
possess a spin degree of freedom, the model could give rise to a U(1)
gauge theory~\cite{Konopka:2008hp}.
In \cite{Konopka:2008ds},
Konopka has analytically and numerically studied the above graph model
as a statistical model.
A phase transition was found,
where it was argued that the low-temperature
phase can be related to spacetime only if the graph can interact with
some matter degrees of
freedom. In~\cite{Hamma:2009xb,Caravelli:2011kq}, a related model,
which in
addition to graphs corresponding to spacetimes, also incorporates a
matter field that resides on the vertices, was proposed to study the
role of matter in the emergence of spacetime from graphs. In
\cite{Conrady:2010qz},
Conrady has constructed a Hamiltonian favoring low-temperature,
two-dimensional manifolds through terms that explicitly favor
two-dimensional triangulations; for example, each vertex is favored to have 6 edges as in
a triangular lattice, and tetrahedra are penalized. The model was simulated for small system sizes
($N\leq 180$ edges), which showed a heat capacity peak, and a transition
temperature that decreased with system size.

In this paper, we also investigate a statistical model of graphs, in
that the objects under consideration are merely abstract graphs, without
any information on the positions of the vertices, or the lengths of
the edges. A graph can randomly transform into another graph according
to a set of transformation rules. Graphs with given numbers of vertices and edges are considered, and they are governed by a
Hamiltonian that favors graphs with a set of local symmetries. If these local symmetries are preserved, the resulting graphs should be nearly triangulations of manifolds with a certain dimensionality, where the dimensionality is controlled by the ratio of vertices to edges. We are interested in whether any global structure of the graphs arises as a consequence.

Because every edge in this model corresponds to a positive length
$\epsilon$, only real positive distances can arise, so this model can
only be used to describe Riemannian manifolds (i.e., with positive
definite metric). The metric of a Riemannian manifold can be
alternatively viewed as a distance function between any pair of
points, which satisfies the triangle inequality. On a graph, there is
also a natural notion of distance, namely the length of the
shortest path between a pair of vertices. This distance is also
positive-definite and satisfies the triangle inequality. Thus on any graph, there is
a well-defined distance function, as well as a
corresponding geodesic. Graph geodesics between two
vertices are often highly degenerate, however, unlike the case for
manifolds. If a manifold is to emerge from a graph, one expects that
in the continuous limit all degenerate
geodesics are close by, and the differences of their paths are only
of order $\epsilon$. After establishing this distance function between vertices, mapping the graph to a Riemannian manifold is still a non-trivial problem. If we enforce that every edge is identical in that they have the same length when mapped to the Riemannian manifold, then only for certain graph configurations will a Riemannian manifold emerge from the graph. Otherwise the system will be frustrated and unable to meet the condition of constant edge length, without increasing the dimension above that of the manifold that would emerge from the graph.

In this paper, after reviewing the relevant graph theory
preliminaries, we introduce a graph Hamiltonian based only upon local
symmetries. We evolve the graph under the Monte Carlo rules obeying
statistical mechanical equilibrium, and we investigate whether a
low-temperature manifold state emerges. We investigate the sharpness
of the phase
transition using energy as an order parameter for different size
systems, and we discuss the likely first-order nature of the
transition. We construct heat capacity curves as a function of
temperature and investigate the transition temperature as a function
of system size, which points toward a zero-temperature phase
transition in the bulk limit. The Haussdorf dimensionality of the
emergent manifold is investigated and found to be an increasing
function of system size, and approximately 3 for the largest system
sizes we investigated (2000 vertices). Correlation functions between
defect-carrying vertices and edges are investigated to determine
whether the effective potential between defects is attractive or
repulsive. Finally, we argue in analogy to condensed-matter systems
that a nonzero phase transition temperature requires long-range
interactions, and show that a Coulombic-like term between graph
vertices yields an apparently finite-phase transition temperature, but
with a highly ramified manifold.

\section{Graph theory preliminaries}

\begin{figure}
\centering
\includegraphics{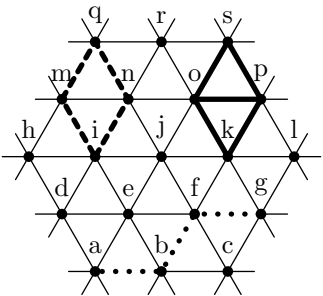}
\caption{\label{FigDefs}Examples for the graph theory concepts.}
\end{figure}

Before motivating for details of the model, we shall remind the reader about some graph theory concepts, which will be needed later in constructing the model.

A \textbf{graph} $G$ is composed of a set of vertices $V(G)$ and a set of edges $E(G)$, where every edge is a subset of $V(G)$ with two elements. Note that by this definition, the two vertices in an edge set cannot be the same vertex, and two edges cannot connect the same two vertices. Such graphs are sometimes called ``simple graphs,'' as opposed to ``multigraphs.'' Because we will only consider graphs of this definition, they will be simply referred to as ``graphs.''

A vertex $v$ is \textbf{incident} with an edge $e$ if $v\in e$. We denote an edge $e$ by its vertices, or \textbf{ends}, say $u$ and $v$, as $e=\{u,v\}$, or simply $e=uv$. A vertex $u$ is a \textbf{neighbor} of, or is \textbf{adjacent} to, a vertex $v$ if $uv$ is an edge. The \textbf{valency} or \textbf{degree} of a vertex is the number of edges incident to that vertex.

A graph in which every vertex has the same valency is \textbf{regular}. It is $k$-regular if every vertex has valency $k$.

A graph in which every pair of vertices is connected by an edge is \textbf{complete}. It is denoted by $K_n$ if it has $n$ vertices.

$G'$ is a \textbf{subgraph} of a graph $G$, if $G'$ is a graph, $V(G')\subseteq V(G)$ and $E(G')\subseteq E(G)$, and this is denoted by $G'\subseteq G$.

If $U\subseteq V(G)$, the subgraph $G'$ \textbf{induced} by $U$ is the graph for which $V(G')=U$, and $E(G')$ contains an edge $xy$ if and only if $x,y\in U$ and $xy\in E(G)$. This is denoted by $G'=G[U]$, and $G'$ is called an \textbf{induced subgraph} of $G$.
(For example, in Fig.~\ref{FigDefs}, the vertices $k,o,p,s$, and the five thick edges, compose an induced subgraph; the vertices $i,m,n,q$, and the four thick dotted edges, compose a subgraph, but not an induced subgraph.)
In particular, in a graph $G$, the subgraph induced by the set of neighbors of a vertex $v$ is called the \textbf{neighborhood} of $v$, and is denoted by $G_N(v)$.

A \textbf{path} is an alternating sequence of vertices and edges, beginning with a vertex and ending with a vertex, where each vertex is incident to both the edge that precedes it and the edge that follows it in the sequence, and where the vertices that precede and follow an edge are the end vertices of that edge. The \textbf{length} of a path is the number of edges in the path. (For example, in Fig.~\ref{FigDefs}, $(a,ab,b,bf,f,fg,g)$ is a path with length 3, in which the edges are denoted by dotted lines, and is also one of several paths between $a$ and $g$ having the minimal distance.) The \textbf{distance} between two vertices is the length of shortest path between them. In a graph $G$, the distance between vertices $u,v$ is denoted by $d_G(u,v)$.

A graph is \textbf{connected} if any two vertices are linked by a path.

The \textbf{eccentricity} $\epsilon_G(v)$ of a vertex $v$ in a graph $G$ is the maximum distance from $v$ to any other vertex, i.e.,
\sen{
\epsilon_G(v)=\max_{u\in V(G)}d_G(v,u),
}
where $d_G(v,u)$ is the distance between $v$ and $u$ in the graph $G$.

\begin{figure}
\includegraphics[scale=1.0]{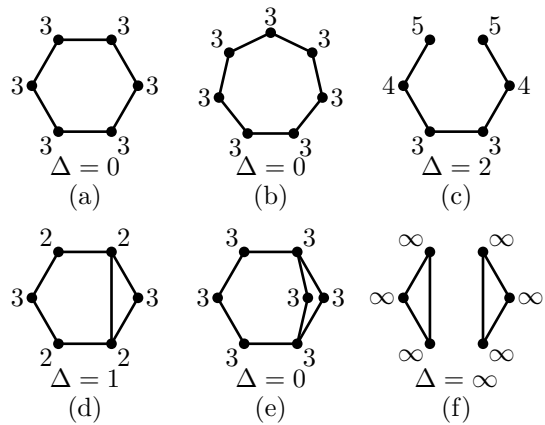}
\caption{\label{FigNbh}Some examples of neighborhood subgraphs. The eccentricity is labeled for each vertex, and the difference of diameter and radius of these subgraphs, which is denoted by $\Delta$, is labeled below each graph. (a) is the neighborhood subgraph of vertices in the triangular lattice graph (Fig.~\ref{FigDefs}); (b), (c), (d) and (e) appear commonly in simulations, as parts of the defects; (f) is the neighborhood subgraph of vertices in the graph in Fig.~\ref{FigExotic}.}
\end{figure}

The \textbf{diameter} $\textrm{diam}(G)$ of a graph $G$ is the maximum eccentricity over all vertices in a graph, and the \textbf{radius} $\textrm{rad}(G)$ is the minimum,
\sen{
\textrm{diam}(G)=\max_{v\in V(G)}\epsilon_G(v),\quad \textrm{rad}(G)=\min_{v\in V(G)}\epsilon_G(v).
}
When $G$ is not connected, $\textrm{diam}(G)$ and $\textrm{rad}(G)$ are defined to be infinite. Some examples of neighborhood subgraphs are shown in Fig.~\ref{FigNbh}. For every vertex in Fig.~\ref{FigDefs}, the neighborhood subgraph is Fig.~\ref{FigNbh}(a); for every vertex in Fig.~\ref{FigExotic}, the neighborhood subgraph is Fig.~\ref{FigNbh}(f). Figures \ref{FigNbh}(b)-\ref{FigNbh}(e) are examples of neighborhood subgraphs that appear commonly in the simulation.

Given a lattice, the corresponding \textbf{lattice graph} is the graph whose vertices are the points in the lattice, and whose edges are the pairs of nearest points in the lattice. (For example, the whole graph in Fig.~\ref{FigDefs} is an equilateral triangular lattice graph.)

\section{The model}

To gain intuition on the form of constraints and Hamiltonians that may
induce manifolds, let us construct some graphs resembling some
manifolds, starting with the example of a flat two-dimensional plane
$\R^2$.
Intuitively, any two-dimensional lattice graph as defined above forms a ``two-dimensional'' manifold, and a coordinate system of the manifold naturally inherits from the coordinates of the lattice
graph. This is directly analogous to a Bravais lattice in crystallography.
{\it A priori} there seems no decisive reason to choose any particular
Bravais lattice as the preferred graph configuration; however, we shall choose the
equilateral triangular lattice graph (Fig.~\ref{FigDefs}), using the
following argument. On $\R^2$, for any point $p$ and any distance
$\delta$, let $B_\delta(p)$ denote the geodesic ball centered at $p$
with radius $\delta$, and $B_\delta(p)-p$ has the topology of a circle
$S^1$. For graphs, we can define the notion of ``geodesic ball''
similarly with that in Riemannian geometry. Let $B_n(v)$ be the set of
the vertices that have distance from vertex $v$ no greater than $n$,
including $v$ itself. For any two-dimensional lattice, if we denote
the corresponding graph by $G$, for sufficiently large $n$, the
induced subgraph $G[B_n(v)-v]$ also looks like $S^1$
topologically. However, for $n=1$, namely the
neighborhood subgraph $G_N(v)=G[B_1(v)-v]$, this property is no longer
true for all lattices. For example, on the square lattice, $G_N(v)$ is composed of 4
disconnected vertices. Only for the equilateral triangular lattice,
$G_N(v)$ looks topologically like $S^1$. Thus in this sense,
the equilateral triangular lattice graph is the closest analog to $\R^2$
among all the two-dimensional lattice graphs, on all distance
scales down to~$\epsilon$.

A graph can form a two-dimensional lattice for the correct ratio of edges to
vertices.
While a thermalized lattice in two dimensions is isotropic \cite{Mermin:1966fe,Hohenberg:1967zz,Coleman:1973ci}, the connectivity of such a lattice is
still well-defined at low temperature. We thus choose to add defects in the
form of extra edges or bonds, which will evolve under some
Hamiltonian.  This allows bonded vertices
to be permuted, so that the low-temperature phase is still a
``quasi-fluid'' that retains a symmetry corresponding to randomized
graph connectivities. The extra edges induce defects in the lattice,
which may be mobile.
The exact shape of the defects and the reason why the
defects are unstable or meta-stable depend sensitively on the Hamiltonian. We
shall construct a candidate Hamiltonian, and test the stability of the
defects by numerical simulation. This construction generalizes
to $\R^n$ straightforwardly: We can see that the defect-free lattice is the $n$-dimensional lattice as arising from a regular tiling of $n$-dimensional tetrahedra.
The defect is a $(n-1)$-dimensional ``foam'' that divides the space into many patches
of lattices with random orientations.

We seek the simplest Hamiltonian that can give rise to manifold-like
triangulation graphs as classical solutions, which contain defects
that facilitate graph permutation symmetry. We assume that the action is local, in the sense that it should be a sum over the vertices and/or edges, such that each term involves a finite number of vertices and/or edges within some cutoff distance. This condition is imposed because almost all physics models for which the Hamiltonian or Lagrangian is an integral of the corresponding density are local in the same sense.

A defect manifests itself as a local structure containing vertices with anomalous valency. One obvious local property of manifold-like graphs is that all
vertices not in any defects would have the same valency. Moreover it is likely
that vertices in the defects have just one more or one less
neighbor.
These properties can be enforced by a Hamiltonian quadratic
in the valency:
\se{
H_1=c_1\sum_{v\in V(G)}n_v^2,
\label{EqnH1}
}
where $n_v$ is the valency of vertex $v$, and $c_1$ is a positive
constant (which will be taken to be infinite as described below).
The average valency of the vertices is given by
\se{
\alpha=\frac{2N_E}{N_V}
\label{eqnalpha}
}
where $N_E$ is the total number of edges and $N_V$ the total number of
vertices. Note that, for example, $\alpha=6$ is compatible with a
regular equilateral triangular lattice, which in turn implies that the
emergent manifold is two-dimensional, while $\alpha=12$ is compatible
with the face-centered cubic lattice, which implies a three-dimensional emergent manifold. Thus without changing the form of the
Hamiltonian, we should be able to find manifolds with different
dimensionalities by adopting different {\it a priori} values of
$\alpha$. In the simulations described below, we choose $\alpha$ to be
a non-integer, so that there exists an
``excess'' number of edges, which contribute to the presence of
defects. Because the total number of vertices and edges are fixed, the
term in (\ref{EqnH1}) is minimized when every vertex has
valency either $\lfloor\alpha\rfloor$ or $\lceil\alpha\rceil$. In our
simulations, $c_1$ is taken to be infinite and so is no longer an
adjustable parameter of the model, and the corresponding term in
\eqref{EqnH1} is enforced to be minimal.

\begin{figure}
\includegraphics{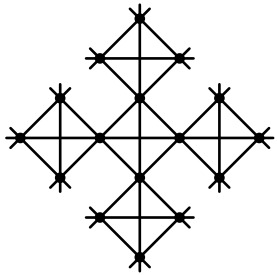}
\caption{\label{FigExotic}A 6-regular graph that is not similar to any manifold. This graph can be viewed as an infinite rooted ``tree'' graph, in which each node has three children (except the root node has four children), and every node of the ``tree'' is actually a tetrahedron.}
\end{figure}

To obtain manifold-like solutions consisting of patches of
close-packed lattices interspersed with defects, it is not sufficient to
impose only the condition that each vertex has
approximately the same number of neighbors. Many regular graphs do not
look like any manifold at all (see, for example,
Fig.~\ref{FigExotic}). Additional terms in the Hamiltonian are
thus required for manifold-like solutions.

One candidate for such a term consists of particular
subgraphs that can be embedded into the graph. From this
viewpoint, $n_v$ is the number of $K_2$ subgraphs (two vertices
connected by an edge) that go through the vertex $v$.
It is likely however that choosing more terms of this type will affect
the dimensionality of the resulting spacetime. For example, if we
incorporate terms that favor more $K_3$ subgraphs (triangles) and fewer
$K_4$ subgraphs (tetrahedra), then it can be expected that these terms
would favor two-dimensional manifolds \cite{Conrady:2010qz}. As we hope to find a model that
does not select the dimensionality at the level of the Hamiltonian, we
will not use any other term of this type besides $H_1$.

Another property of manifold-like graphs is that around most
vertices, the graph has a local discrete rotational symmetry that reflects the local isotropy of the emergent manifold. This can be restated as for each vertex $v$, the subgraphs $G[B_n(v)-v]$ for most $v$ should have a discrete rotational symmetry. To reduce the number of
possible Hamiltonian terms, we impose this condition only on $G[B_1(v)-v]$, which is also
$G_N(v)$. We introduce the term
\se{
H_2=c_2\sum_{v\in V(G)}\Delta(v),
\label{EqnH2}
}
where $c_2$ is a positive constant, and
\se{
\Delta(v)=\textrm{diam}(G_N(v))-\textrm{rad}(G_N(v)),
}
in which $\textrm{diam}(G_N(v))$ is the
diameter of the subgraph $G_N(v)$, and $\textrm{rad}(G_N(v))$ is the
radius of the subgraph $G_N(v)$. By the definitions of diameter and
radius of graphs, if the subgraph $G_N(v)$ is not connected, they are
both infinite. Here, we additionally define that their difference
$\textrm{diam}(G_N(v))-\textrm{rad}(G_N(v))$ is also infinite when
$G_N(v)$ is not connected. The term $H_2$ then enforces that all
neighborhood subgraphs are always connected. When $G_N(v)$ is
connected, the difference between its diameter and its radius is a
measure of its asymmetry. Figure~\ref{FigNbh} shows several examples of neighborhood subgraphs. The eccentricity of every
vertex in the subgraphs is labeled, along with the value of $\Delta(v)$ for each subgraph. For Figs.~\ref{FigNbh}(a) and (b), the $G_N(v)$'s have a rotation symmetry of $\mathbb{D}_6$ and $\mathbb{D}_7$, respectively, while Figs.~\ref{FigNbh}(c)-(e) are not rotationally symmetric.

In two dimensions, a graph forms a triangulation of a surface if and
only if all the neighborhood subgraphs are
cycles~\cite{Malnic1992147}. When the degrees of the subgraphs are
either 6
or 7, which is imposed by the $H_1$ term, one can see from the
examples in Fig.~\ref{FigNbh} that the $H_2$ term indeed favors
cyclic neighborhood subgraphs, with only one exception shown in Fig.~\ref{FigNbh}(e). We thus expect that, in this model, a graph with low
energy is almost a triangulation of a surface.

Thus we propose the following model: Consider a simple graph with
$N_V$ vertices and $N_E$ edges. All the vertices are
labeled, so isomorphic configurations with different labeling are
considered to be different configurations. The Hamiltonian is composed of two terms, as motivated previously:
\se{
H=H_1+H_2.
\label{EqnH}
}

Because the Hamiltonian is prohibitive to analytical solution, we implement
a numerical simulation, as described in the next section, to study the equilibrium
states of this model in the canonical ensemble, i.e. at a given
temperature. In particular,
we will be interested in the structures of the states with low
energies, and the nature of the phase transition, if one exists, to
these low-energy states.

\section{Numerical simulation}

We sample equilibrium states in the model using a Monte Carlo
simulation \cite{Newman:1999}. The parameter $\alpha$ defined in
\eqref{eqnalpha} as giving the mean number of edges per
vertex is taken to be slightly larger than 6, which we expect will
induce two-dimensional structures dictated by triangulations as described
above. There is no
fixed boundary on the graphs. The size of
the graphs is specified by the number of vertices $N_V$, and the
number of edges $N_E$. For convenience, in the following we use $N_V$
and the number of extra edges $X \equiv N_E-3N_V$, to specify the size of the
graphs. Given the graph size, the initial configuration is taken to be
a randomly generated, connected graph.

The graph is evolved in the canonical ensemble with temperature
$1/\beta$. In each Monte Carlo step, one end of an edge can jump from
one vertex to another. We randomly pick an edge, and randomly label
its ends by $u$ and $v$. To find the new location of the edge $uv$,
we perform a random walk starting from $v$ as the origin, which does not pass through the
edge $uv$ (this condition guarantees that a connected graph remains
  connected after such a move).  The number of steps $\ell$ of the
walk is a random positive integer chosen from the probability distribution
$P(\ell)=\gamma^{\ell-1}-\gamma^{\ell}$, where $\gamma$ is a parameter
between 0 and 1 (we take $\gamma=0.5$ below). Denote the ending vertex of the random walk as
$v'$. The edge is then moved from $uv$ to $uv'$. If the new graph is
still simple, its energy is compared with that of the old graph, and
this move is accepted or rejected according to the Metropolis
algorithm \cite{Newman:1999}. Each ``sweep'' through the system contains $N_E$ Monte Carlo
steps, so on average each edge has one chance to jump in one sweep.
Such a method is ergodic;  moreover with this jumping
scheme, the energy of only a few vertices
is affected after each Monte Carlo step, and the energy of only these vertices needs to be updated.

Simulations are performed with $c_1=\infty$, $c_2=1.0$, $\gamma=0.5$,
and various values of $N_V$, $X$ and $\beta$. Before showing the thermodynamics results from the simulations, let us first describe the method that we used to render a graph from the simulations, in order to interpret its evolution.

\subsection{\label{SecRendering}Rendering graphs}

To render a graph such that its structure can be best visualized, we need to devise an appropriate drawing scheme. A drawing of a graph maps vertices to points in $\R^n$ with
line segments connecting adjacent points. The following method is used
to generate drawings in $\R^3$. For any drawing of a graph $G$, we seek
to minimize the function
\se{
H_{\textrm{draw}}=\sum_{e\in
  E(G)}\left(a_1l_e^2+\frac{a_2}{l_e^2}\right)+\sum_{\substack{i,j\in
    V(G),\ i\ne j,\\i,j\textrm{ not adjacent}}}\frac{a_3}{l_{ij}^2},
\label{EqnHDraw}
}
where $l_e$ is the length of the drawing of edge $e$, $l_{ij}$ is the
distance of the drawing between vertices $i,j$, and
$a_1=1.0,a_2=1.0,a_3=5.0$. The first term gives a preferred length for
every edge, and the second term gives a repelling force to every
non-adjacent pair of vertices. The function $H_{\textrm{draw}}$ is
chosen this way in order to make every edge have approximately the
same length in the drawing, and as well, to make the drawing as
expanded as possible. In practice, even for moderate-sized graphs,
$H_{\textrm{draw}}$ has numerous local minima and is difficult to
minimize. We thus use another Monte Carlo calculation to search for
its near-optimal values. Initially, all the vertices are located at
the origin of $\R^3$. In each Monte Carlo step, a randomly-chosen
vertex is randomly moved to another position within the ball of radius
$\delta=2.5$, centered at the original position, and the new position
has uniform probability distribution within the ball. After the Monte
Carlo calculation, because the low-temperature configurations in the
model are conjectured to be similar to triangulations of surfaces, we
also search for all the $K_3$ subgraphs (triangles) in the graph, and
render (flat, solid) triangles to fill the interior of the $K_3$'s.

\begin{figure*}
\includegraphics[width=14cm]{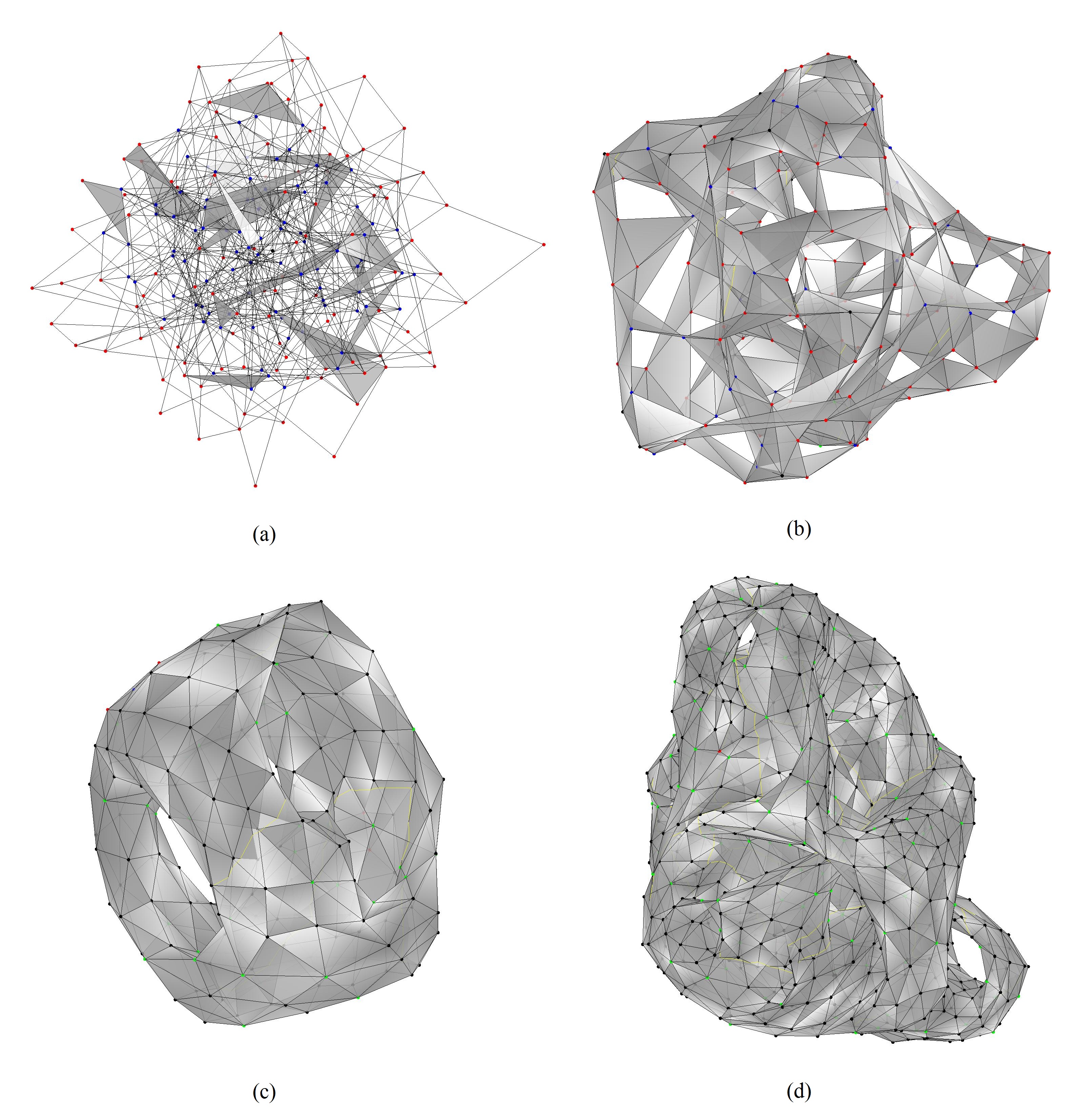}
\caption{\label{FigSample}(Color online) Some snapshots from the simulations, drawn in three
  dimensions. Panels (a)-(c) are for the system with number of
  vertices $N_V=200$ and number of extra edges $X=20$,
  where (a) is the initial configuration, (b) is a typical
  configuration at high temperature ($\beta=1.0$), and (c) is a
  typical configuration at low temperature ($\beta=2.0$). Compared
  with the sphere, the drawing (c) has three more handles, and the
  surface intersects with itself in three places, so it has a
  non-trivial, non-orientable topology. Panel (d) is for the system of size
  $N_V=1000$ and $X=100$, and shows a typical configuration at low
  temperature ($\beta=2.0$). In these drawings, if a vertex has
  valency 6, it is black if its $\Delta$ value is zero, and is red if its
  $\Delta$ value is nonzero; if a vertex has valency 7, it is green if its
  $\Delta$ value is zero, and is blue if its $\Delta$ value is nonzero (see
  text). As well,  yellow lines are drawn at places where two
  triangles intersect, and the manifold thus passes through itself.}
\end{figure*}

Figure~\ref{FigSample}
shows some snapshots taken from the simulations. Figures~\ref{FigSample}(a)\hll{-(c)} are for the system of size
$N_V=200$ and $X=20$. Figure~\ref{FigSample}(a) shows the initial
configuration, \ref{FigSample}(b) shows a typical configuration at
high temperature ($\beta=1.0$), and \ref{FigSample}(c) shows a typical
configuration at low temperature ($\beta=2.0$). Figure~\ref{FigSample}(d) is for the system of size $N_V=1000$ and $X=100$,
and it is a typical configuration at low temperature ($\beta=2.0$).

In the sample drawings in Fig.~\ref{FigSample}, different colors are
used to denote different types of vertices. The color-code is as follows:
\begin{center}
\begin{tabular}{c|c|c}
\hline
& Degree$=6$ & Degree$=7$\\
\hline
Zero contribution to $H_2$ & black & green\\
Nonzero contribution to $H_2$ & red & blue\\
\hline
\end{tabular}
\end{center}
Also, yellow lines are drawn at places where two triangles intersect,
i.e., this identifies where the triangulated surface intersects with
itself.

\begin{figure*}
\includegraphics[width=14cm]{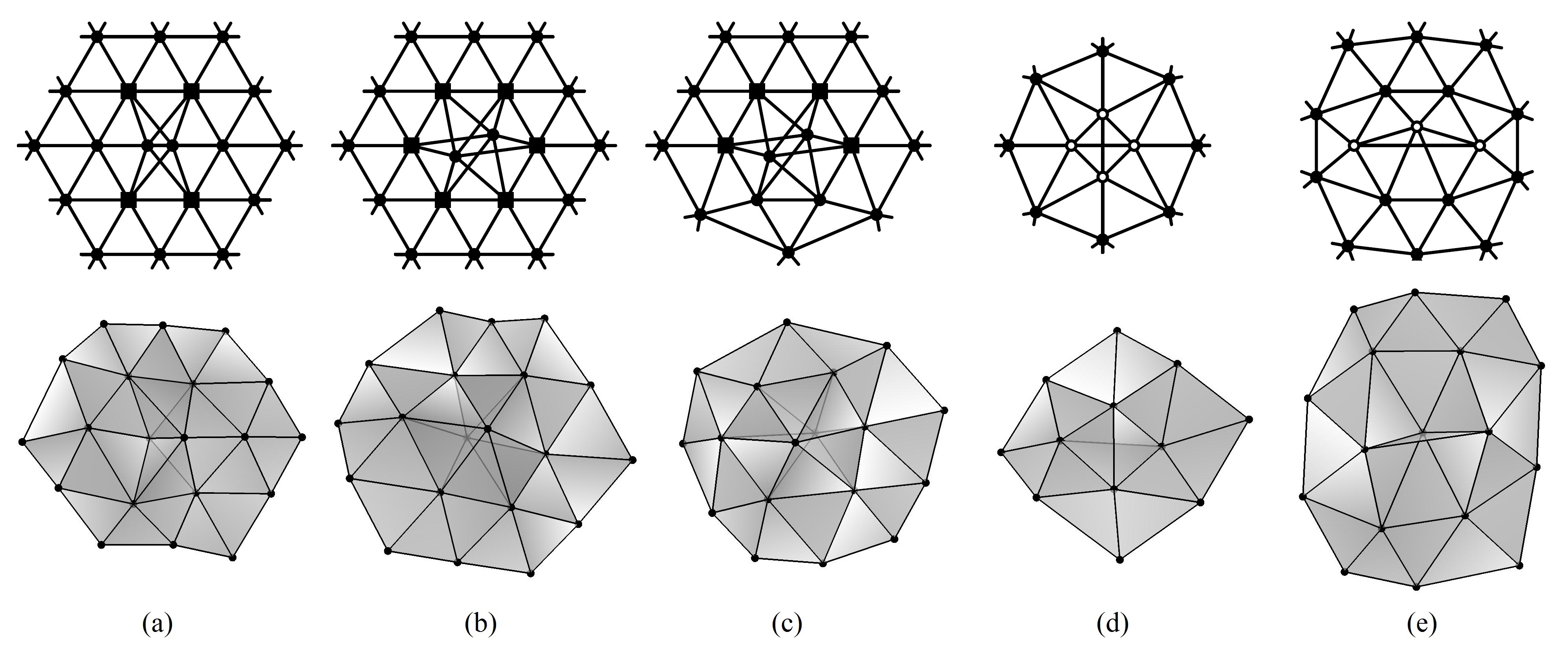}\\
\caption{\label{FigDefect}Examples of some common defects. \hll{Once
    the graph is triangulated to construct a surface, defects
  (a-d) have ``bubble-wrap'' morphology, while defect (e) has
  ``frenulum'' morphology.} The figures in the first row are
  the schematic drawing of the defects, in which a vertex is marked with a
  square if its valency is 7, a vertex is marked with an open circle
  if it contributes positive energy to $H_2$, and otherwise a vertex
  is marked with a filled circle. The figures in the second row are the
  corresponding drawings of the defects using the method described in
  subsection~\ref{SecRendering}.
  Compared with the equilateral
  triangular lattice, examples (a), (b), (c), (d) and (e) have 2, 3,
  2, 0, 0 extra edges, respectively.}
\end{figure*}

\subsection{Topology of the manifold in the presence of defects}

For the low-temperature graphs, several examples of common local
defects are shown in Fig.~\ref{FigDefect}. They are called local in
the sense that in the vicinities of these defects, the graph is
similar to some triangulation of surfaces with trivial
topology. Among these examples, \hll{the ``bubble-wrap'' defects}
(a)-(c) do not increase the total
energy, and around such defects the ratio between the number of edges
and vertices is larger than 3. In other words, these defects can
``absorb'' the extra edges without energy cost. Also note that (a) and
(b) do not change the long range order of the lattice orientation,
while (c) does alter the long range order. Taken together, these
defects induce configurational degeneracies in all the energy levels,
including the ground state energy level, and at the same time induce
graph permutation symmetry by randomly breaking the lattice's long range order, at
least in the rendering scheme of the manifold described
above. \stt{Defects} \hll{The bubble wrap defect}
(d) and \hll{``frenulum'' defect} (e) increase the total energy, and alter the lattice
orientation more drastically.

As discussed above, low-temperature graphs in the model are
similar to two-dimensional triangulated surfaces. However, they
contain local defects, and there are overall topological features of
the surfaces that emerge from the graphs. For example, in the drawing
Fig.~\ref{FigSample}(c), one can see that the emergent surface
contains several handles, and the surface intersects itself in several
places. In the drawing Fig.~\ref{FigSample}(d), the
topology of the emergent surface is too intricate to easily identify. The
Hamiltonian does not constrain the topology in any way, so in
general, emergent surfaces of low-temperature graphs in the model have
complicated topologies. The emergent surfaces have potentially many
handles, and are in general non-orientable, in that there is no
separation between interior and exterior sides of the surface. In our
simulations, we also observe that the topology of the graphs' emergent
surfaces can dynamically change, even at a low energy.

We note, however, that the
choice of $N_V$ and $N_E$ can constrain the topology. At low
temperatures, the graphs are nearly
triangulations, albeit with potentially complicated topologies. If a graph is strictly a triangulation, and we denote the number of triangles as $N_F$, then the Euler
characteristic $\chi$ of the surface is given by
$\chi=N_V-N_E+N_F$. For a triangulation, $3N_F=2N_E$; and we
previously defined $N_E=3N_V+X$. Putting these three equations
together, we find $\chi=-X/3$. As we showed above, defects on the
graphs can absorb edges, so the relation for the
nearly-triangulated graphs becomes an inequality
$\chi\ge-X/3$. In addition, for any surface, $\chi\le2$, with
  $\chi=2$ corresponding to the topology of a sphere.
Thus the Euler characteristic $\chi$ of the emergent surface can take
any integer value between $-X/3$ and $2$. The $X$ values used in our
simulations are not very small, so this constraint still allows for
many possible different topologies for the emergent surface.

\subsection{Phase transition}

In this sub-section we study the transition between the low-/high-
temperature phases. For system sizes $N_V=100,200,300,500,1000,1500,2000$,
and number of excess edges $X=0.1N_V$, the expectation value of energy
$\langle E\rangle$, and the heat capacity $C=\beta^2\left(\langle
  E^2\rangle-\langle E\rangle^2\right)$ are computed for various
inverse temperatures $\beta$, where the angle bracket here means averaging
over all the samples in a simulation.

\begin{figure}
\centering
\includegraphics{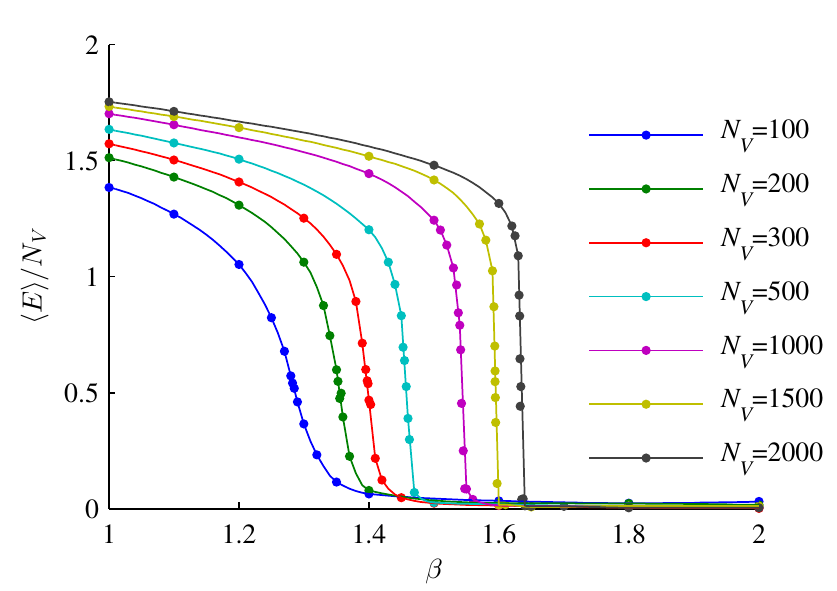}
\caption{\label{FigChart1}(Color online) The average energy density $\braket{E}/N_V$ as a function of inverse temperature $\beta$ for for several $N_V$'s indicated in the legend.}
\end{figure}

\begin{figure}
\centering
\includegraphics{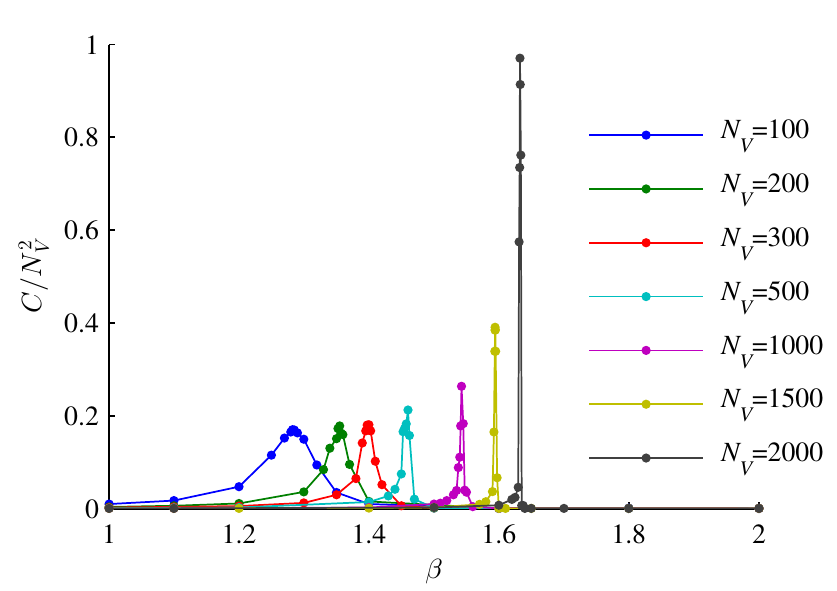}
\caption{\label{FigChart2}(Color online) The rescaled heat capacity $C/N_V^2=\beta^2\left(\braket{E^2}-\braket{E}^2\right)/N_V^2$ as a function of inverse temperature $\beta$ for several $N_V$'s indicated in the legend.}

\end{figure}

\begin{figure}
\centering
\includegraphics{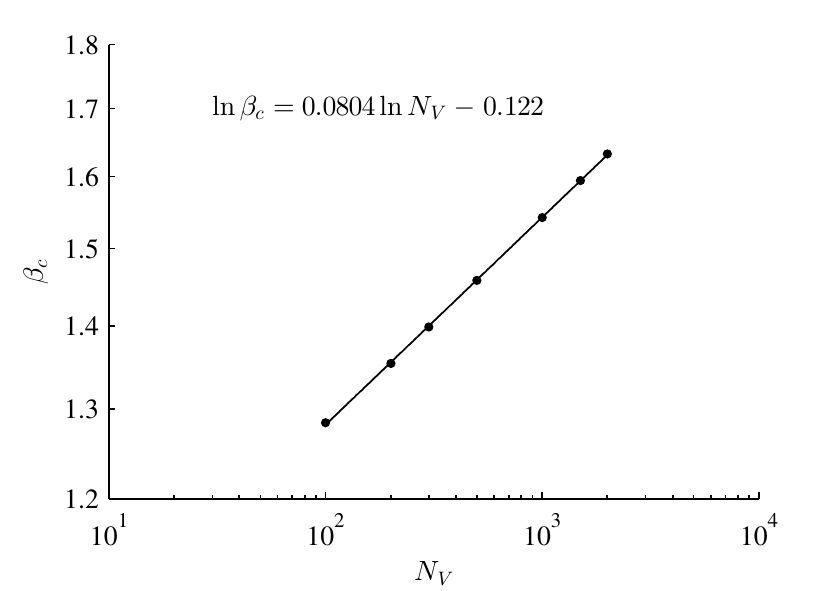}
\caption{\label{FigLogLog}Log-log plot of the inverse transition temperature $\beta_c$ in the model as a function of system size $N_V$, and the best fit line. The straight line fit indicates that as $N_V\rightarrow\infty$, the transition temperature $T_c\rightarrow0$.}
\end{figure}

\begin{figure}
\centering
\includegraphics{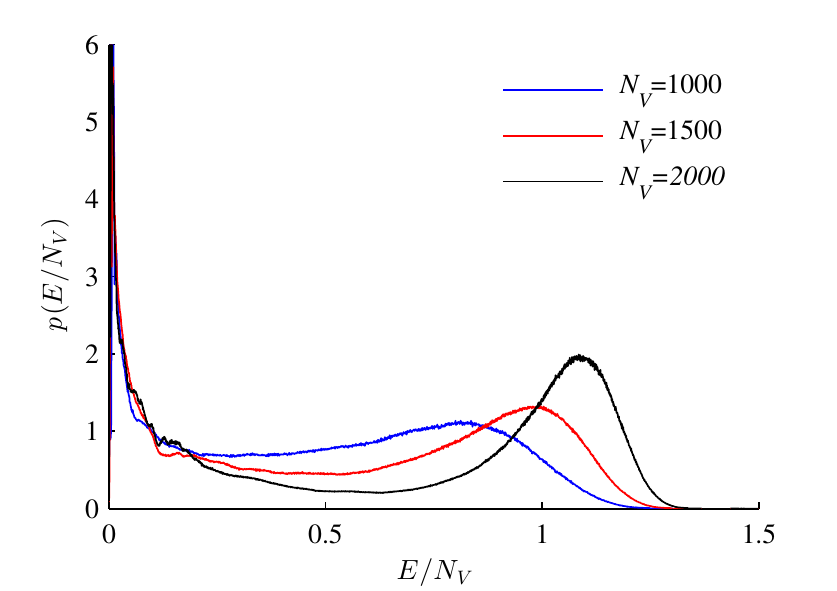}
\caption{\label{FigHisto}(Color online) The probability density of the intensive energy $E/N_V$ for the
  systems of size $N_V=1000$, $1500$ and $2000$, at each system's
  transition temperature. The error of $p(E/N_V)$ for $E/N_V\ge0.5$ is
  small ($\Delta p \leq 0.1$), the error for $0.01<E/N_V<0.5$ is
  $\Delta p \leq
  0.6$, and the error for the smallest values of energy $E/N_V\le0.01$ is $\Delta p \leq 2.5$.}
\end{figure}

The results are shown in Fig.~\ref{FigChart1} and Fig.~\ref{FigChart2}. For the three largest systems with $N_V=1000$, $N_V=1500$, and
$N_V=2000$, we also employ the weighted histogram analysis
method (WHAM)\footnote{See Appendix A for the details of our implementation of WHAM.}
\cite{Ferrenberg:1988yz} to improve the sampling quality. The
inverse transition temperature $\beta_c$ is defined as the inverse
temperature where the heat capacity is maximal. It can be seen that
$\beta_c$ increases as $N_V$ increases, \hll{an effect also seen
  previously in
  other graph models} \hlref{\cite{Konopka:2008ds,Conrady:2010qz}}
Near the transition
temperature $\beta_c$, \hll{$|d\braket{E}/d\beta|$} also increases as $N_V$
increases, and thus the widths of the heat capacity peaks
decrease as $N_V$ increases, indicating the transition becomes more
cooperative.
Figure~\ref{FigLogLog} shows a log-log plot of the inverse transition
temperature as a function of $N_V$. The linear relation in the plot
indicates that as $N_V$ goes to infinity, the transition temperature
would go to zero. In addition, Fig.~\ref{FigHisto} shows the
probability density distribution of $E/N_V$, for the systems of size
$N_V=1000$, $1500$, and $2000$, at each system's transition
temperature. As $N_V$ increases, the energy distribution of the two
phases become more bimodal, \hll{and the temperature-dependence of
  the heat capacity  in} \hlref{Fig.~\ref{FigChart1}} \hll{becomes
  sharper, indicating a more cooperative transition with increasing
system
size}\hlref{~\cite{PlotkinSS02:quartrev1,PlotkinSS02:quartrev2}}.
\hll{Together this implies}
that the transition is first order in the bulk limit, with a
corresponding nucleation barrier\hlref{~\cite{Binder1987theory}.} \hll{That is, a Landau functional using system
energy as an effective order parameter has a double-well structure with
corresponding barrier separating the low- and high-energy phases}\hlref{~\cite{PlischkeM}.}

In appendix \ref{SecAcceptanceRatio}, we give the acceptance ratio in
  our simulations as a function of inverse temperature. Although the
  acceptance ratio substantially decreases in the low-energy phase,
  the system is still able to undergo dynamics because some local
  defects cost little or no energy.

A transition temperature of zero for infinitely large graphs is
actually not very surprising on entropic grounds. Consider a first order phase transition
of an extended physics model. Denote the size of the system by $N$,
and denote the number of states in the high- and low-temperature phases
by $\Omega_\textrm{H}$ and $\Omega_\textrm{L}$, respectively. Because
the energy difference between these two phases is proportional to $N$,
the phase transition temperature $T_c$ is given approximately by
$\Omega_\textrm{H}e^{-\kappa N/T_c}=\Omega_\textrm{L}$, where $\kappa$
is a positive number. As $N$ increases, for a ``typical'' physics system with short-ranged interactions,
the ratio between $\Omega_\textrm{H}$ and $\Omega_\textrm{L}$ increases
as $e^{\gamma N}$, where $\gamma$ is a positive number. This behavior
results in a finite, nonzero transition temperature in the infinite
size limit. On the other hand, the number of inequivalent graphs with
$N_V$ vertices is typically $N_V^{\gamma' N_V}$, (for example, see
\hlref{\cite{Erdos:1959,Bender:1978,Konopka:2008ds}}) where $\gamma'$ is a positive
number that depends on the constraints of
the allowed graphs. In our case, the allowed graphs should have every
vertex valency equal to six or seven, and every vertex neighborhood
should be connected. While we do not have an algorithm to count the
exact number of allowed graphs, it is reasonable to assume for our
system that the ratio
between $\Omega_\textrm{H}$ and $\Omega_\textrm{L}$ has the typical
asymptotic behavior of graphs, which explains a transition temperature
of zero, i.e., the transition temperature $T_c$ is given by
$N_V^{\gamma'N_V}e^{-\kappa N_V/T_c} \approx 1$.

\begin{figure}
\centering
\includegraphics{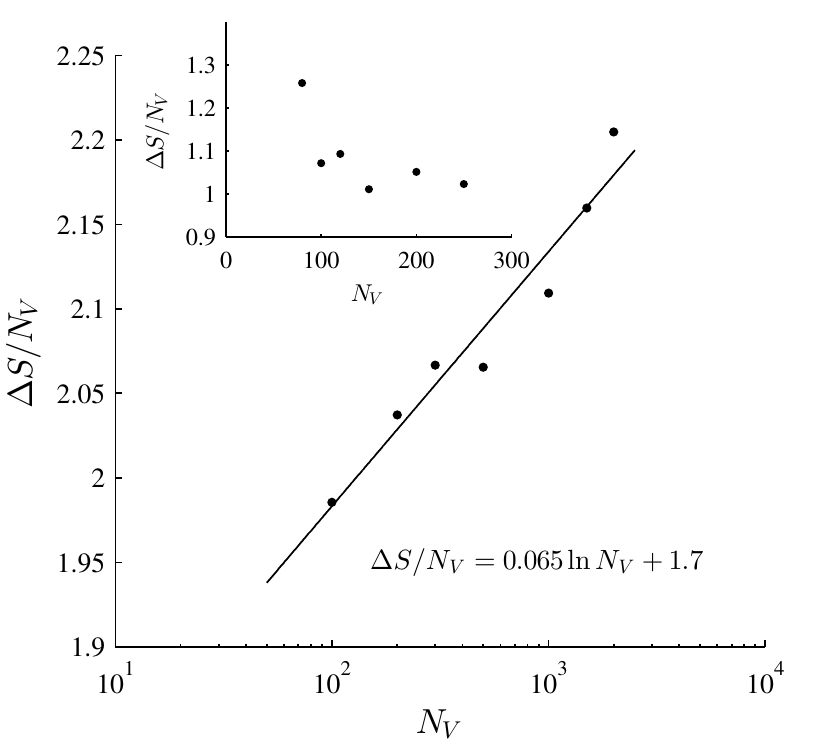}
\caption{\label{FigEntropy}The entropy density difference across the transition $\Delta S/N_V$ as a
  function of $N_V$. The best fit line using a logarithmic function is
  also shown. The inset shows $\Delta S/N_V$ as a function of $N_V$ for a model including a Coulomb potential between valency-7 vertices (see section \ref{sectCoulomb}). Including long-range interactions can remove super-extensivity of the entropy.
}
\end{figure}

To validate the above argument, we can calculate the entropy difference across the transition as given by
\se{
\Delta S=\int_{T_2}^{T_1}\frac{C(T)}{T}dT=\int_{\beta_1}^{\beta_2}\frac{C(\beta)}{\beta}d\beta,
\label{EqnDS}
}
where $C$ is the heat capacity, and $\beta_1$ and $\beta_2$ are
typical inverse temperatures in the high-temperature phase and low-temperature phase, respectively, which are taken to be
$\beta_1=\beta_c-100/N_V$, $\beta_2=\beta_c+100/N_V$, i.e., we ensure that
the window defining the transition narrows as the width of the heat
capacity peak narrows. Fig~\ref{FigEntropy} shows the difference in entropy density $\Delta S/N_V$ as a function of $N_V$, which, rather than remaining constant, is a monotonically increasing function. Thus the entropy of the system is super-extensive. If the ratio $\Omega_\textrm{H}/\Omega_\textrm{L}$ of the model scales like $N_V^{\gamma'N_V}$ as argued above, $\Delta S/N_V$ will have the form $\Delta S/N_V=\gamma'\ln N_V+b$. The best fit line using this logarithmic function is also shown in Fig~\ref{FigEntropy}, which is consistent with a super-extensive entropy, with $\gamma'\simeq0.065$.

\subsection{Geometric properties}

In this sub-section, we analyze some geometric properties of the two phases: if a geometric property is distinct in the two phases, it can serve as an order parameter that signals the phase transition.

\begin{figure}
\centering
\includegraphics{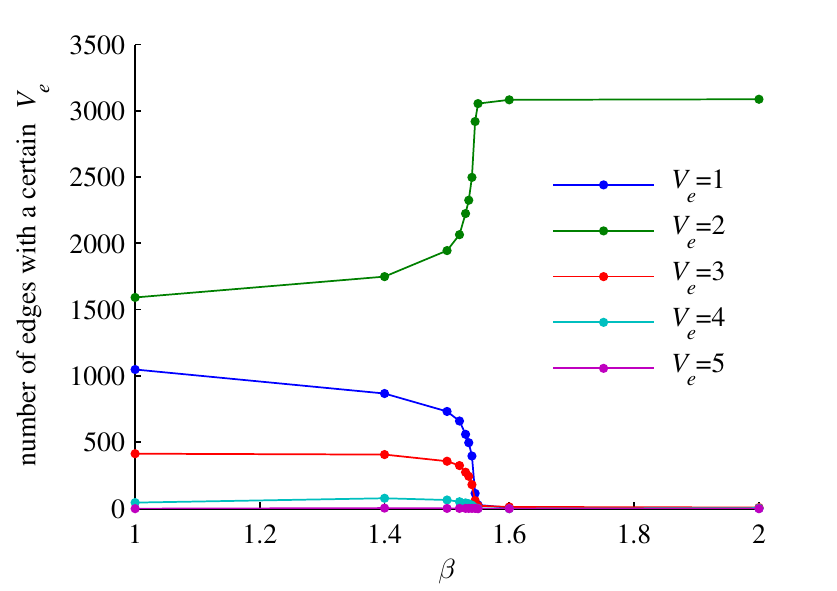}
\caption{\label{FigEdgeValency}(Color online) Distribution of edge valencies as a function of inverse
  temperature $\beta_c$, for the system of size $N_V=1000$. There are
  no edges in the simulation with edge valency less than one or larger than five.}
\end{figure}

As was mentioned before, because the low-energy graphs are nearly
triangulations for our Hamiltonian, it is useful to introduce an order
parameter that measures how similar
graphs are to triangulations. For this purpose we can study the
distribution of edge
valencies, where the edge valency is defined as the number of triangles that an edge is
part of. In a perfect triangulation of a surface without boundaries, the edge
valencies are always two, so we expect that at low temperatures, the
distribution of edge valency should approximate a delta function
around two.
The distribution of edge
valencies for the system of size $N_V=1000, X=100$ is shown in
Fig.~\ref{FigEdgeValency} as a function of temperature. Indeed, almost all edges have edge
valency two at temperatures below the transition temperature. Near the
transition temperature however, there is a sudden change in the distribution of
edge valencies: above the transition temperature, edge
valencies both above and less than two appear.

\begin{figure}
\centering
\includegraphics{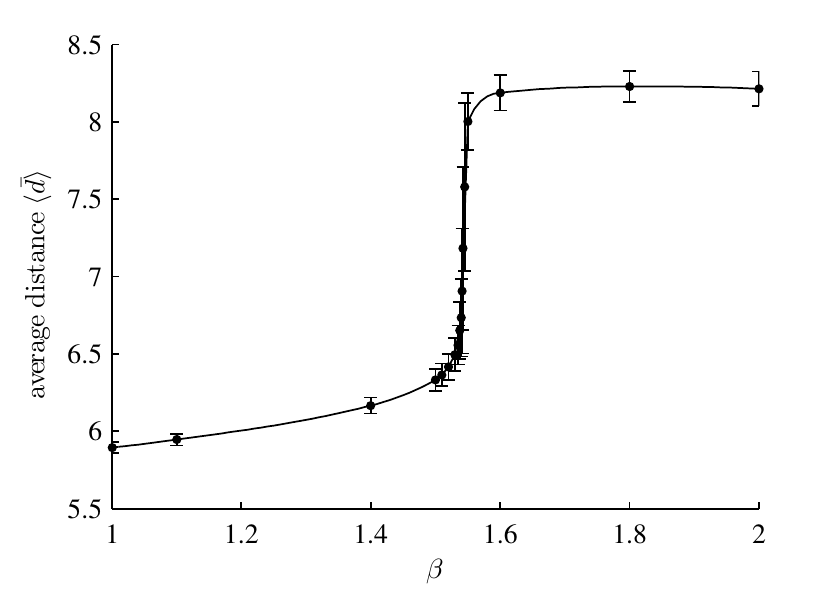}
\caption{\label{FigAvgDist}Average distance $\braket{\bar d}$ between pairs of vertices, plotted as a function of inverse temperature $\beta$, for the system of size $N_V=1000$. $\braket{\bar d}$ is first averaged over all pairs of vertices in a given snapshot, and then averaged over all snapshots at a given temperature. The vertical bars at each data point indicate the standard deviation between snapshots: $\sqrt{ \braket{\bar d^2} - \braket{\bar d}^2}$.
}
\end{figure}

Another quantity that is useful as an order parameter is the average distance between all pairs of
vertices, denoted by $\braket{\bar d}$, where the bar means averaging
over all pairs of vertices in a graph, and the angle bracket means
averaging over samples of an equilibrium simulation. We expect that above the phase
transition temperature, graphs will exhibit ``small world'' topologies and
thus $\braket{\bar d}$ will be relativity small.
The quantity $\braket{\bar d}$ gives the
characteristic linear size of the graphs. Figure~\ref{FigAvgDist}
plots $\braket{\bar d}$ {\it vs.} inverse temperature $\beta$, for
$N_V=1000$. Indeed, the low-temperature phase has a larger $\braket{\bar d}$
than the high-temperature phase; low-temperature
  graphs tend to have much more structure than high-temperature
  graphs, resulting in larger values of $\braket{\bar d}$.

\begin{figure}
\includegraphics{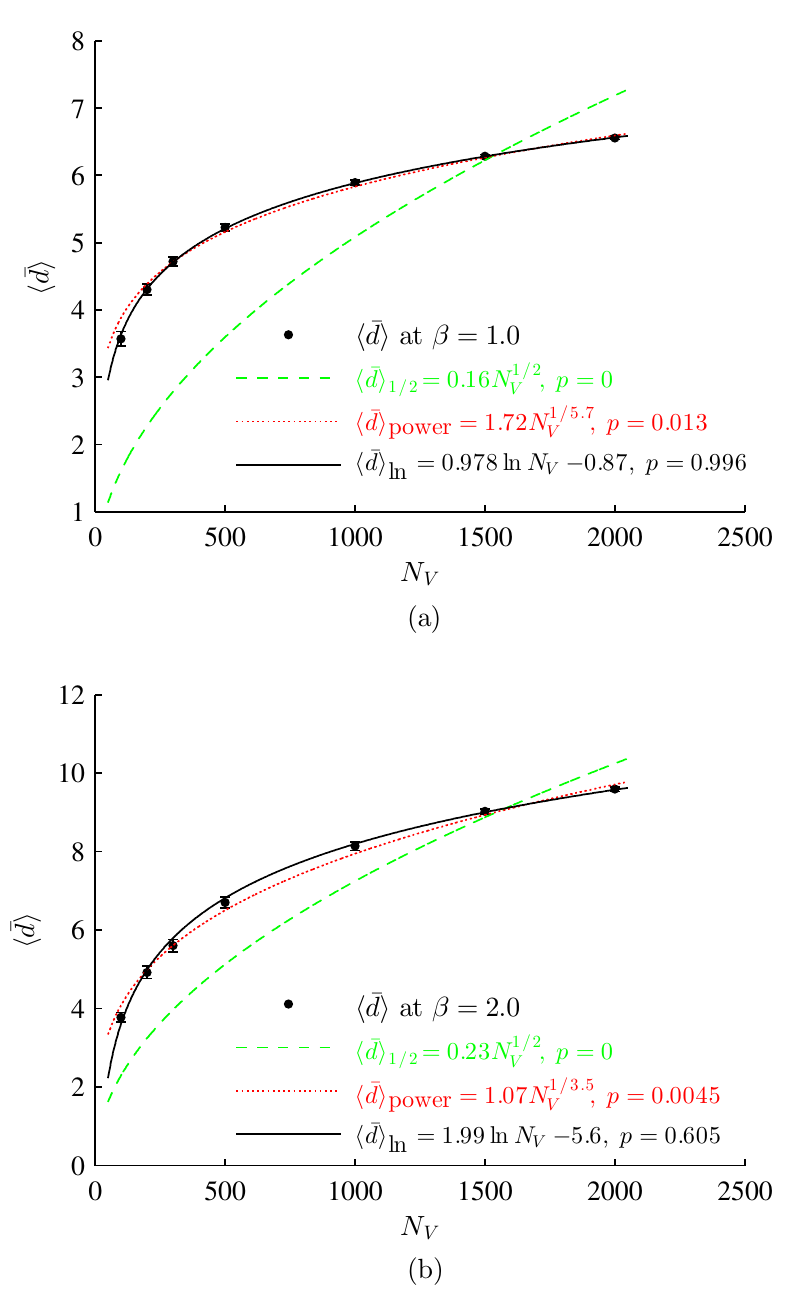}
\caption{\label{FigScaling}(Color online) The average distance
  $\braket{\bar d}$ between pairs of vertices as a function of the system size $N_V$ (discrete
  points), and the best fit lines using a square root function (green dashed lines), using
  a power function (red solid lines), and using a logarithmic function (black solid
  lines). Plots are shown both above the transition ($\beta=1.0$) in panel (a), and below the transition ($\beta=2.0$) in panel (b). For each best fit line, its expression and $p$-value are also shown, where the $p$-values are calculated for the null hypothesis that the residues
    $(d_{\textrm{fit}}-\braket{\bar d})/\delta d$ come from a normal
    distribution with variance smaller than 1. For both temperatures,
  the logarithmic function gives the best fit to the measured
  data.}
\end{figure}

In Fig.~\ref{FigScaling}, the average distance $\braket{\bar d}$ is
shown as a function of the system size $N_V$, at $\beta=1.0$ (above
the transition) and at
$\beta=2.0$ (below the transition). The best fit lines using a logarithmic function and using
a power function are also shown in Fig.~\ref{FigScaling}. The $p$-value for each best fit line is calculated for the null hypothesis that the residues $(d_{\textrm{fit}}-\braket{\bar d})/\delta d$ come from a normal distribution with variance smaller than 1, so that a higher $p$-value indicates a better model. These
relations between $\braket{\bar d}$ and $N_V$ can be understood by
comparing with random graphs, which generally display ``small-world''
connectivity, with average distances growing logarithmically with the
number of vertices \cite{Erdos:1959}.
In our model, the Hamiltonian only constrains the graphs locally, so these graphs satisfy small-world
behavior in the high temperature phase accurately, as shown by the logarithmic best fit line in
Fig.~\ref{FigScaling}(a).
For the low temperature phase, we can define an effective scaling dimension (see, e.g., \cite{Creswick:1992})
\se{
D_s=\frac{d\ln N_V}{d\ln\braket{\bar d}}.
\label{eqDs}
}
On a non-fractal surface, $\braket{\bar d} \sim N_v^{1/2}$, i.e., $D_s=2$. However, it
is seen from Fig.~\ref{FigScaling}(b) that the residuals with the square root function are too
large. If we take $D_s$ as a parameter in the fitting, a power-law function with $D_s\simeq3.5$ is a much better fit to the empirical scaling.
Perhaps surprisingly however, the logarithmic function is still the best fit function, indicating that the low-temperature graphs still display
small-world connectivity.
Enforcing a power-law fit at every system size, i.e., $\braket{\bar d} \sim N_v^{1/D_s\left(N_v\right)}$, would induce a variable dimensionality in the exponent.

\begin{figure}
\centering
\includegraphics{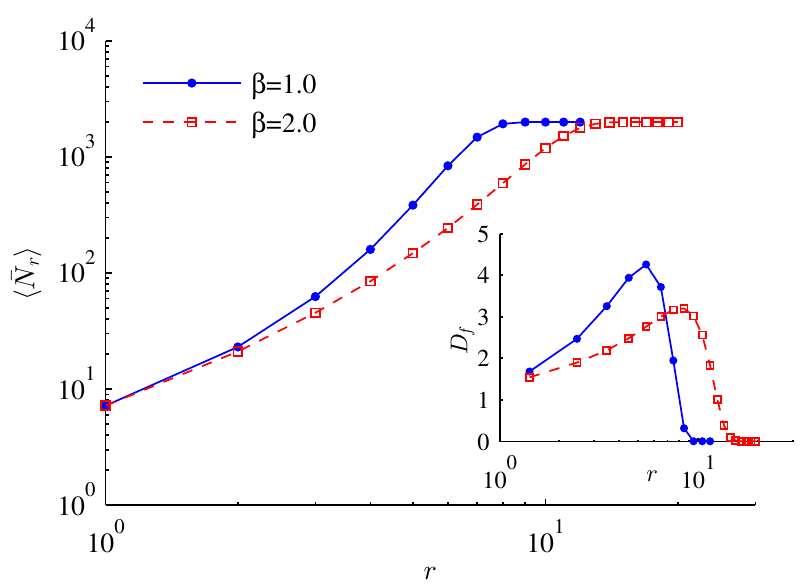}
\caption{\label{FigNrLogLog}(Color online) Log-log plot of $\braket{\bar N_r}$, the thermally averaged number of
vertices within a distance $r$, as a function of $r$; the slope gives
the dimensionality of the system, which in this case is
distance-dependent. The plot shown is for the system with size
$N_V=2000$, at $\beta=1.0$ (blue line) and at $\beta=2.0$ (red line), which
bracket the transition. For the same system, the inset shows the fractal dimension as a function of $r$.}
\end{figure}

Another related definition of dimensionality measures the increase in
number of vertices with distance from a given vertex. On a graph, one
can pick an arbitrary central vertex, and count how many vertices
$N_r$ have distance no greater than $r$ from that center.
We can then average both over all central vertices and over all
equilibrium configurations at a given temperature, denoting the doubly
averaged volume by $\braket{\bar N_r}$. If
$\braket{\bar N_r}$ increases with $r$ polynomially, the
fractal (Haussdorf) dimension can be defined as
\se{
D_f=\frac{d\ln\braket{\bar N_r}}{d\ln r}.
\label{EqnFractalDim}
}
In practice the dimension of the graph may itself depend on the radius $r$, so it makes sense
to talk rigorously about the dimensionality of a graph only if $D_f$ is essentially
constant over some range of $r$. A log-log plot of $\braket{\bar N_r}$
vs. $r$ is shown in Fig.~\ref{FigNrLogLog}, for $N_V=2000$ at
$\beta=1.0$ and $\beta=2.0$, where the slope thus gives
the dimensionality and is shown in the inset. One can see that the effective dimension $D_f$ is smaller below the transition.
Consistent with the previous analysis using \eqref{eqDs}, there
is no well-defined dimension for the graphs, which are
small-world-like. Instead there is an increasing dimensionality with
increasing length scale, until boundary effects of the system are
felt. The dimensionality has values around 2 for small values of $r$, because of the
local lattice-like structure; it is
also small for very large values of $r$, because a finite-sized graph
must eventually be bounded, at which point
$\braket{\bar N_r}$ will no longer increase polynomially at large $r$. Table~\ref{TabFractalDim} lists the maximal value of $D_f(r)$ for systems with different sizes, at inverse temperatures $\beta=1.0$ and $\beta=2.0$. As the table shows, $D_{f,\textrm{max}}$ increases with $N_V$, which indicates that as $N_V$ increases, there is no universal fractal dimensionality that can be approached by the graphs. Instead, the graphs are still small-world.

\begin{table}[h]
\centering
\begin{tabular}{c|c|c}
\hline
& $\beta=1.0$ & $\beta=2.0$ \\
\hline
$N_V=1000$ & $3.62$ & $2.72$ \\
$N_V=1500$ & $3.99$ & $3.11$\\
$N_V=2000$ & $4.26$ & $3.20$ \\
\hline
\end{tabular}
\caption{The maximal value of the fractal dimension $D_f$ as defined in \eqref{EqnFractalDim} for systems with $N_V=1000$, $N_V=1500$ and
  $N_V=2000$, at inverse temperatures $\beta=1.0$ above the transition
  and $\beta=2.0$ below the transition.}
\label{TabFractalDim}
\end{table}

The small-worldness of the low temperature graphs in the bulk limit
can be viewed as a consequence of the graph Hamiltonian in \eqref{EqnH}, which is a sum of
local terms. The defects in the manifold are also local --- in the bulk these have no
effect on the large-scale structure of the resulting graphs. This is
manifested for finite-size graphs by the fact that as $N_V$ increases,
the topologies of graphs become progressively more complicated, see
e.g. Fig.~\ref{FigSample}(c) and Fig.~\ref{FigSample}(d).
The manifolds contain numerous handles and surface
intersections, so that a planar dimensionality does not adequately
describe the system.
In this sense there is already the signature in
the low-temperature phase of the finite system that the bulk system is always
disordered.

\subsection{Correlation functions}

\begin{figure}
\includegraphics{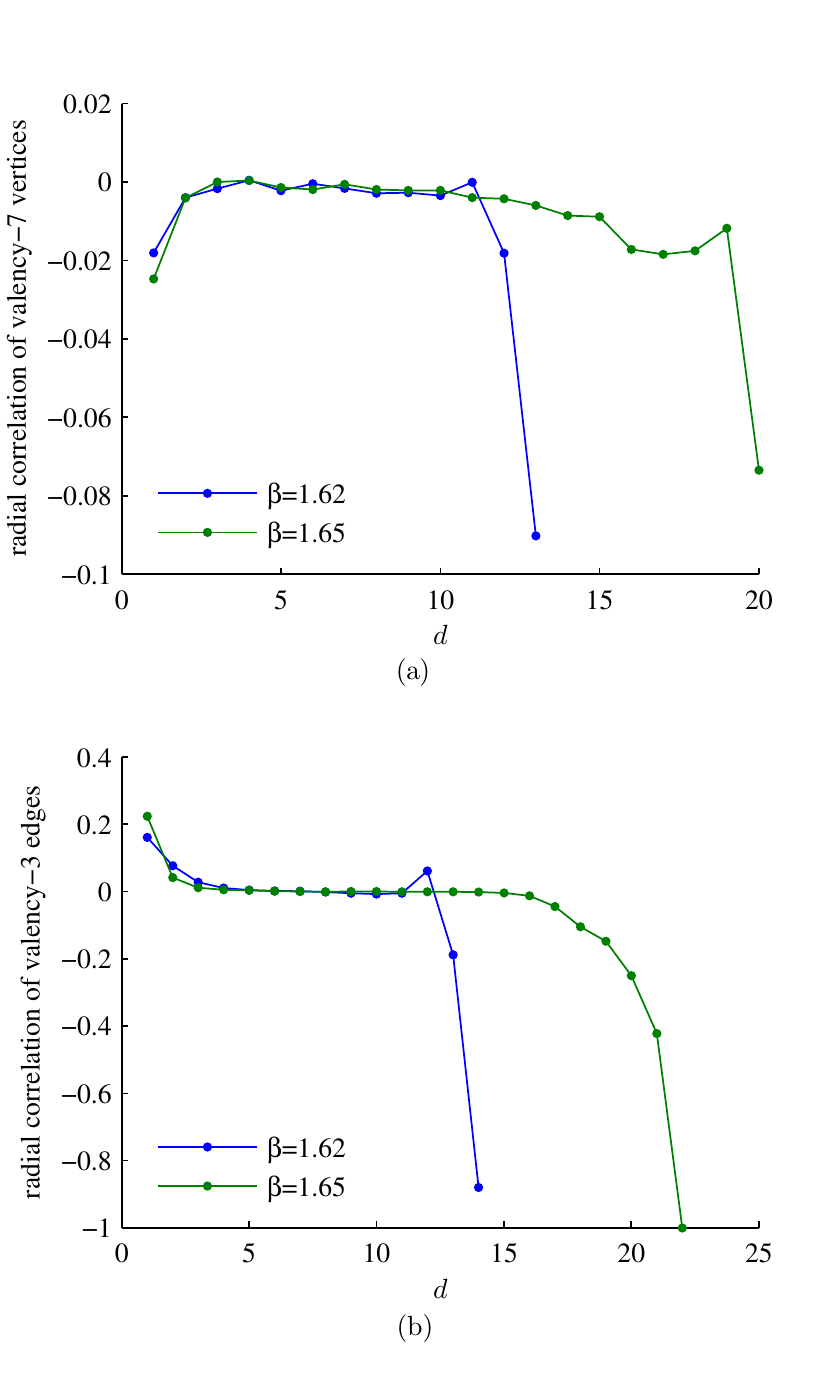}
\caption{\label{FigCorr}(Color online) Radial correlation function defined through
  \eqref{corrf} of (a) valency-7 vertices and (b) valency-3
  edges. Correlations are calculated for the system with size
  $N_V=2000$ at $\beta=1.62$, which is in the high-temperature phase, and at $\beta=1.65$, which is the low-temperature phase.
}
\end{figure}

Defects in this model such as those shown in Fig.~\ref{FigDefect}
contain irregularities that make them differ from part of a regular
lattice. However, regions far away from them may not be affected by
their existence; i.e., there may be no long-range correlation between
such defects. In this subsection, we define and calculate
correlation functions between defect pairs.

Because valency-7 vertices induce defects, we first measure the radial
correlation function of valency-7 vertices. In general, the
correlation between two random variables $X,Y$ with expected values
$\mu_X,\mu_Y$ and standard deviations $\sigma_X,\sigma_Y$ is defined as
\se{
\textrm{corr}(X,Y)=\frac{E[(X-\mu_X)(Y-\mu_Y)]}{\sigma_X\sigma_Y},
\label{corrf}
}
where $E$ is the expectation value operator. In our case, we take all
pairs of vertices with distance $d$ in a graph; $X$ is 1 if the first
vertex in a pair has valency 7, and 0 otherwise, and $Y$ is defined
similarly for the second vertex. Then the correlation function is
averaged over all equilibrium samples. The result for $N_V=2000$,
taken at inverse temperatures $\beta=1.62$ and $\beta=1.65$, which are marginally below and above $\beta_c$ respectively, is shown in
Fig.~\ref{FigCorr}(a). When the distance $d$ is very small ($d=1$ or $2$), the
correlation function deviates from zero, because of the local
structure of of the defects (see Fig.~\ref{FigDefect}), which in this case induces anti-correlation. For intermediate values of $d$ ($3\le d\le 10$), the correlation
is very small, indicating the defects are uncoupled. However, for large values of $d$, the correlation
function becomes negative. This is because the valency of a vertex,
and the distance from this vertex to other vertices, are not
independent: compared with the valency-6 vertices, the valency-7
vertices tend to have smaller distances to other vertices.
For example, for $N_V=2000, \beta=1.62$, the mean distance to valency 6 vertices is 7.18, while the mean
distance to valency 7 vertices is 7.04. Thus it is
less probable to find two valency-7 vertices with a large distance,
and hence they anti-correlate at large distances. The correlation
function is quite small over a range of $d$ as one might anticipate,
but the above global effect makes it difficult to quantitatively
confirm that defects are decoupled at large distance.

As another measure of the correlation between defects, we can measure
the radial correlation function of valency-3 edges, since their
existence indicates deviation of the graph from a triangulation of
surface. For example, every defect in Fig.~\ref{FigDefect} contains
valency-3 edges. The distance between two edges is defined by
taking the 4 vertices defining the two edges, and finding the pair of vertices with the
minimum distance between them.  Since a pair of edges having a common
vertex would then have a distance of zero, we add one to the above
definition of edge distance.
The results for $N_V=2000$, at $\beta=1.62$ and $\beta=1.65$, are shown in Fig.~\ref{FigCorr}(b). At small distances
($d\le3$), there exists short range positive correlation between the
valency-3 edges --- the mean force between them is attractive, due
again to the particular structure within a given low energy local
defect. At large distances ($d\ge14$ for $\beta=1.62$, $d\ge17$ for $\beta=1.65$), the correlation function becomes
negative, because valency-3 edges correlate with valency-7 vertices,
which in turn anti-correlate at large distances for the reasons
described above. However for a wide
range of intermediate distances, this correlation function is also
nearly zero, indicating again that the defect attraction is
short-ranged.

\section{\label{sectCoulomb}Addition of a Coulomb potential}

We found above that as the graph size $N_V$ increased
to infinity, the transition temperature $T_c$ approached zero
(Fig.~\ref{FigLogLog}). This is apparently a universal property of
models based on graphs, due to the super-extensive entropy of the high-temperature random phase. Similar arguments appear in the theory of
phase transitions of low-dimensional systems
\cite{landau1958statistical}, wherein the non-extensive energy cost of defect
formation is outweighed at any nonzero temperature by the (extensive)
free energy due to translational entropic gain, so long as
interactions are sufficiently short-ranged. This analogy motivated us to
introduce a model with long-ranged interactions between defects,
anticipating that in such a defect-filled system incurs super-extensive
energetic cost, which may in turn result in a nonzero transition
temperature.

Thus, in addition to the original two terms in the Hamiltonian
\eqref{EqnH}, we introduce a nonlocal Coulomb potential term
to the
Hamiltonian, which gives a repulsive force between any pair of
degree-7 vertices,
\se{
H_3=c_3\sum_{v,u\in V(G),v\ne u}\frac{\delta_{n_v,7}\delta_{n_u,7}}{d(v,u)}.
\label{EqnH3}
}
This is one of the simplest non-local Hamiltonian terms that one can
add to the original Hamiltonian. The Coulomb force is chosen to be
repulsive, because most of the high-temperature states are ``small
world'', in that they have smaller average distances than those of low
temperature states, so such a Coulomb potential can suppress the
appearance of these ``small world'' graphs. 

\begin{figure}
\includegraphics[width=8cm]{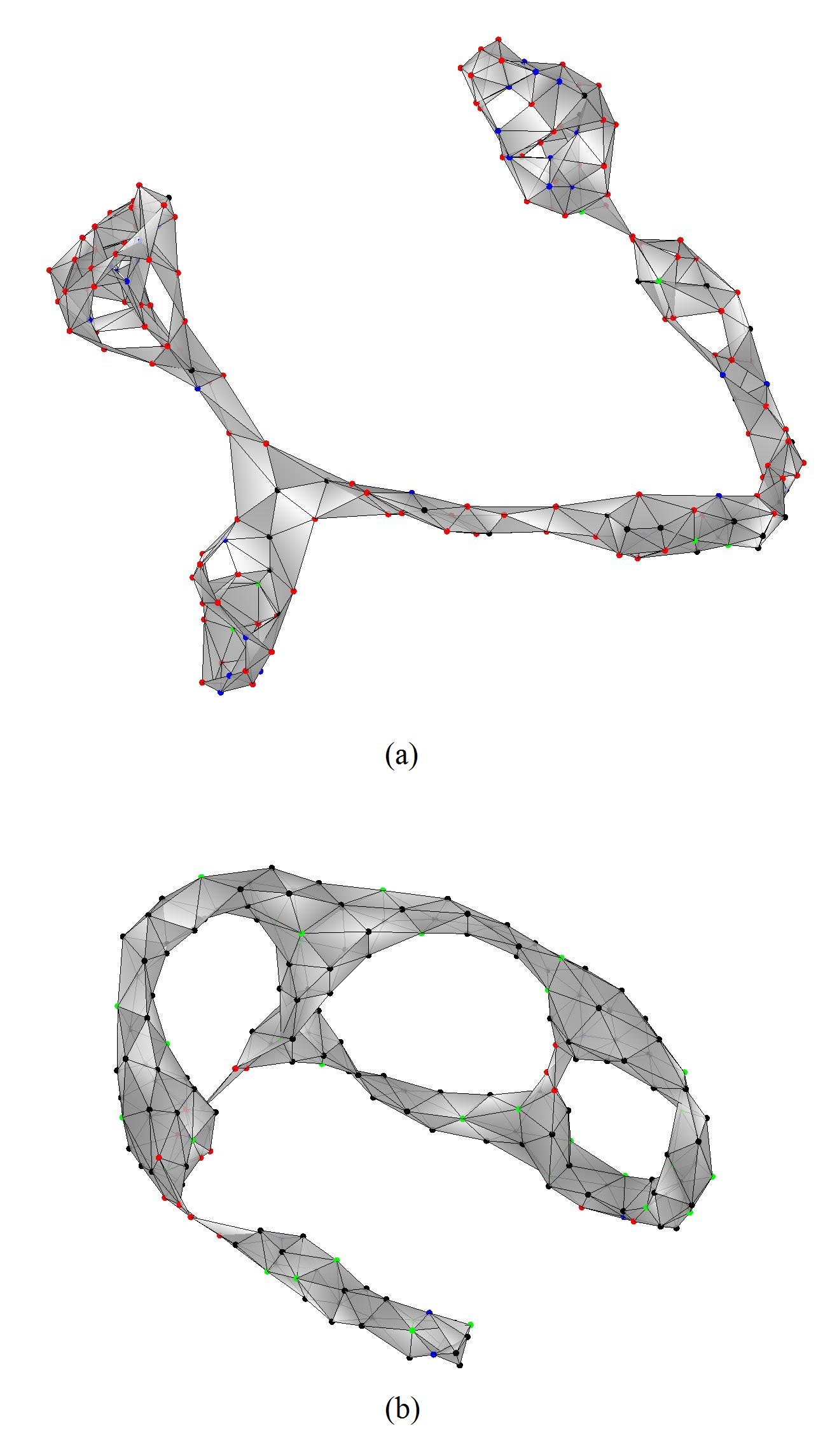}
\caption{\label{FigClbSample}(Color online) Sample configurations for
  the model with Coulomb potential in \eqref{EqnH3} with $c_3=1.0$,
  for the system with number of vertices $N_V=200$ and number of extra
  edges $X=20$,
  drawn in three dimensions. Panel (a) shows a typical configuration
  in the high-temperature phase with $\beta=1.0$; panel (b) shows a
  typical configuration in the low-temperature phase with
  $\beta=2.0$.
}
\end{figure}

\begin{figure}
\includegraphics{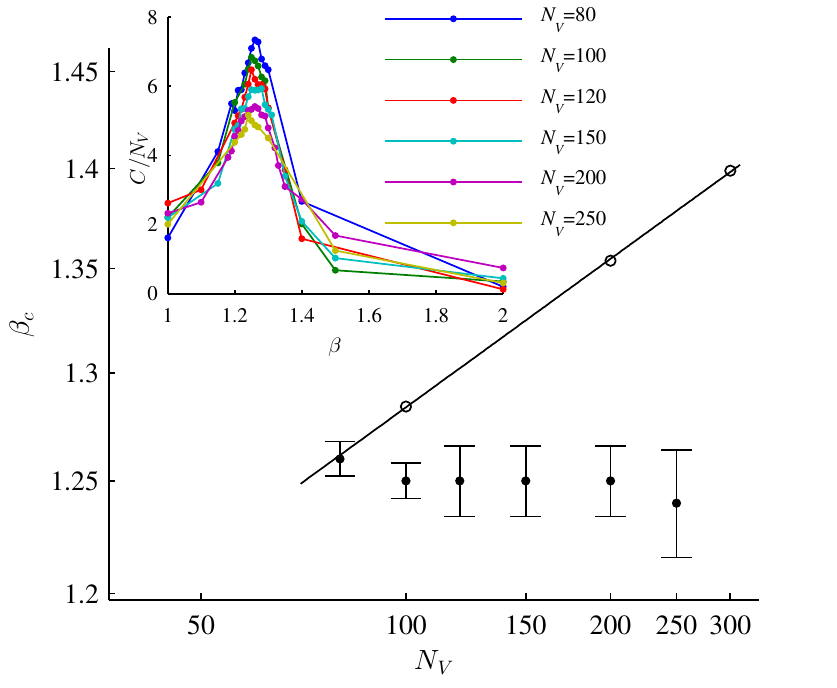}
\caption{\label{FigClbBetac}(Color online) Log-log plot of transition temperatures $\beta_c$ as a
  function of system size $N_V$, for the model with local Hamiltonian
  in \eqref{EqnH} (drawn as circles, with best fit drawn as solid line), and for the model with
  Coulomb potential in \eqref{EqnH3} with $c_3=1.0$ added to
  the local Hamiltonian (discrete
  points with error bars). The inset shows
  the rescaled heat capacity $C/N_V$ as a function of inverse
  temperature $\beta$ for systems with the Coulomb potential added,
  and from which the values and uncertainties of
  $\beta_c$ values are determined.
}
\end{figure}

We test the effect of addition of this Coulomb term by another set of
simulations, in which $c_3=1.0$. Figure~\ref{FigClbSample} shows the
sample drawings of graphs with $N_V=200, X=20$ (a) at high temperature
($\beta=1.0$) and (b) at low temperature ($\beta=2.0$).
These temperatures bracket the heat capacity peak for the system so
that the system is in the disordered and ordered phases respectively
(Fig.~\ref{FigClbBetac}). Because of the
non-locality of $H_3$, simulations are much slower in practice than
before and smaller systems are thus employed:
simulations are performed for $N_V=80,100,120,150,200$ and
$250$, and $X=0.1N_V$. The inset of Fig.~\ref{FigClbBetac} shows the rescaled
heat capacity $C/N_V$ as a function of $\beta$, for several system
sizes. The rescaling factor is now chosen differently than in
Fig.~\ref{FigChart2}, because the systems with the Coulomb
potential have maximal heat capacity approximately proportional to
$N_V$. From the maximal heat capacity, the inverse transition
temperature $\beta_c$ is determined, and is shown in
Fig.~\ref{FigClbBetac} (main panel), in comparison with the
$\beta_c$ values  without the Coulomb potential.

From the graph drawings in Fig.~\ref{FigClbSample}, we can see that because of the repulsive Coulomb
force, both the high-temperature and low-temperature manifold configurations
become rather extended to achieve longer average distances between
defects. This may also explain why the transition temperature does not
change very much with $N_V$:
The characteristic linear size of the systems is much larger when the
repulsive Coulomb potential is present,
which penalizes the increase in complexity that was observed for a
local Hamiltonian as $N_V$ increased. We thus suspect that the entropy
would be extensive for the long-ranged interaction model. To quantify
this, as a final check we plot the entropy change between disordered
and ordered phases as a function of $N_V$ in the inset of Fig.~\ref{FigEntropy}, where $\Delta S$ is calculated by Equation~\eqref{EqnDS}, and $\beta_1=1.0$, $\beta_2=2.0$. As opposed to the entropy difference in the original model, $\Delta S/N_V$ of this model is approximately constant as $N_V$ increases, i.e. the entropy difference is no longer super-extensive, rather it is extensive or sub-extensive.

\begin{figure}
\includegraphics[width=8cm]{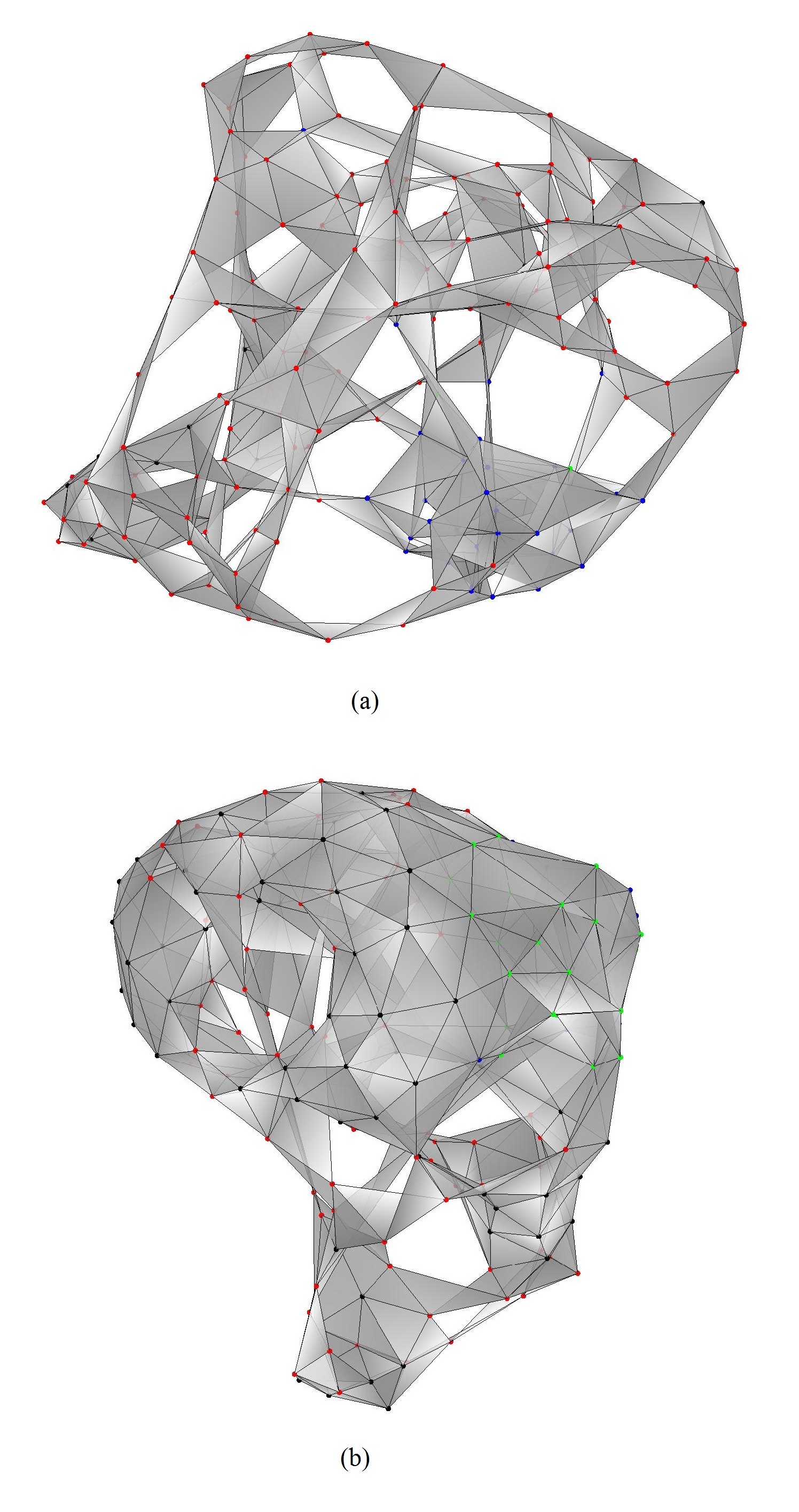}
\caption{\label{FigClbSample34}(Color online) Sample configurations for the model with Coulomb potential in \eqref{EqnH3} with $c_3=-1.0$, for the system of size $N_V=200, X=20$
  drawn in three dimensions. Panel (a) shows a typical configuration
  in the high-temperature phase with $\beta=1.0$; panel (b) shows a
  typical configuration in the low-temperature phase with
  $\beta=2.0$.
}
\end{figure}

We also simulate the model with an attractive Coulomb potential, in
which $c_3=-1.0$. Figure~\ref{FigClbSample34} shows sample drawings of
graphs with $N_V=200, X=20$ (a) at high temperature ($\beta=1.0$) and
(b) at low temperature ($\beta=2.0$). The effect of the attractive
potential can be observed in these samples, in that the valency-7
vertices (green and blue dots) are usually located close together. In
addition, because a local move must involve a valency-7 vertex, the
configuration cannot evolve in the regions composed of purely
valency-6 vertices, and thus the simulation is inefficient. As can be
seen in Fig.~\ref{FigClbSample34}(b), in the region of valency-6
vertices, the configuration does not minimize the Hamiltonian (red
dots have positive contribution to $H_2$), and is not a
triangulation. Thus Fig.~\ref{FigClbSample34}(b) depicts a
long-lived meta-stable state on an energy landscape of
states characteristic of a frustrated system~\cite{PlotkinSS02:quartrev1,PlotkinSS02:quartrev2}. Such a model has
numerous local minima with large
reconfigurational barriers between them, and consequently glassy
relaxation dynamics.

\section{Discussion}

In this paper we have constructed a graph model with a local
Hamiltonian that simply enforces \hll{minimal} valency subject to a
given total number of graph links, along with a graph symmetry between
the local graph radius and diameter.
\hll{The above minimal condition along with fixed total link number gives
rise to near constant valency for all vertices. }
This Hamiltonian gives rise to an
emergent manifold at low temperature.
The one free parameter in the
model does not appear in the Hamiltonian but as an initial condition
of the system. This parameter $\alpha$ determines the edge to vertex
ratio, which is conserved for the system and determines the
dimensionality of the emergent manifold.
When
$\alpha$ is slightly larger than 6, the low temperature solutions have
structural properties consistent with
triangulations
of two-dimensional surfaces.
\hll{We obtained a representation of the emergent manifold by an
  optimization scheme, wherein adjacent vertices are brought as close as
  possible to a certain link distance, non-adjacent vertices are
  repelled from each other, and every triangular subgraph is
  assumed to be filled to render the manifold.}

\hll{The spacetime manifold has historically been treated as a
  triangulation in several previous approaches, in order to regularize
the partition function by constructing discrete analogs to the
continuum manifold}
\hlref{\cite{Weingarten:1982mg,David:1984tx,Ambjorn:1985az,Kazakov:1985ea,Boulatov:1986mm}.}
\hll{For example, in  dynamical
triangulation theory
a given spacetime manifold is triangulated by simplices to calculate
a 
discretized gravitational action }
\hlref{\cite{Johnston:1996rp,Bowick:1997za,Ambjorn:2005qt}.} \hll{ 
In matrix models of gravity, graphs may be constructed as dual to
Feynman diagrams arising from the limit of large internal symmetry
group; by construction the graph constitutes a manifold.
The partition function for two-dimensional quantum gravity can be
expressed as a sum over topologies of triangulated 2D surfaces, for
actions of various forms describing the coupling between matter fields
and spacetime}
\hlref{\cite{DiFrancesco:1993nw};} \hll{
this problem has connections to string theory via the Polyakov action}
\hlref{\cite{Polyakov:1981rd}.}
\hll{The formalism may be extended to study higher dimensional
  generalizations of quantum gravity by group field theory
  models}~\hlref{\cite{Oriti:2006se}.}
\hll{In this context,
  the emergence of a smooth ``hydrodynamic'' spacetime has been described as a condensation of simplicial
  quantum building blocks}~\hlref{\cite{Oriti:2006ar}.}
\hll{Such dual graph triangulations have widely varying vertex valency but
generally represent manifold-like 
surfaces, at least in the condensed phase.
In contrast, the emergent manifolds that we observe have near constant
valency, but often bifurcating morphologies, e.g. the ``bubble-wrap'' or ``frenulum''
defects in Fig.}~\hlref{\ref{FigDefect}.}

\begin{figure}
\includegraphics[width=7.5cm]{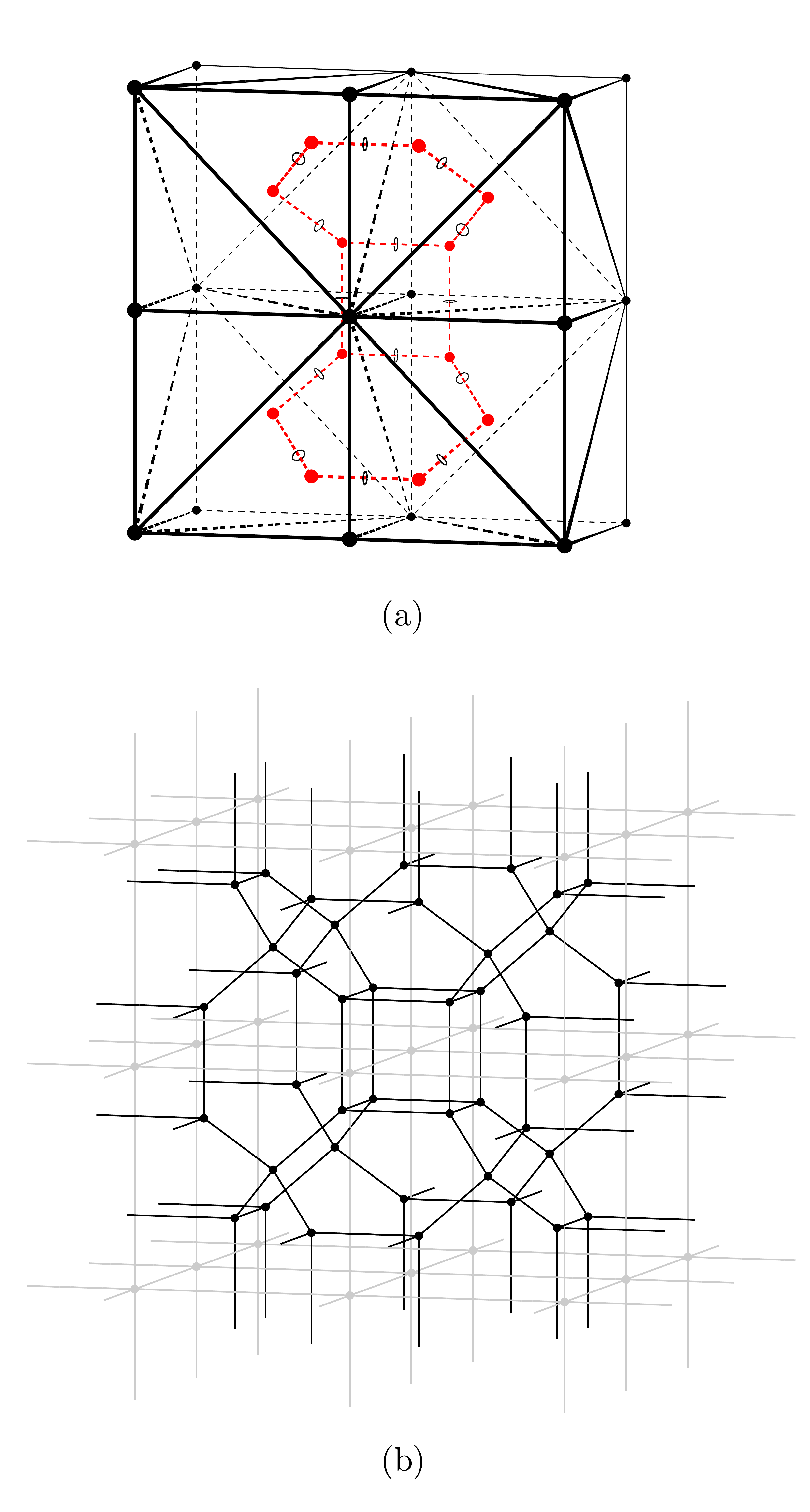}
\caption{\label{FigCubeDual}  \hll{(Color online) (a) A 3D cube triangulated into 5
  tetrahedra}
\hlref{~\cite{ungor2001tiling}}
\hll{may be replicated by translation and
reflection to tesselate the 3D space. Here, part of the dual lattice
is shown as well in red. Red vertices are at the centers of the
tetrahedra in the original triangulation.  At the sites where the dual
lattice bonds pass through the faces of tetrehedra in the original
tesselation, open circles are drawn.
(b) 3D Euclidean space subdivided into the cubes shown in panel (a) (grey
lines); triangulation of the cubes in panel (a) is not shown explictly
here. The thicker black lines correspond to the dual graph of this
triangulation.}
}
\end{figure}

\hll{One can ask whether the present graph model could act as a substitute for the
Feynman diagram construction in matrix models. The Feynman diagram
construction has fixed valency, and is dual to a triangulated
manifold, so a graph model of nearly fixed valency $n_v$ could in principle
give rise to an emergent manifold of dimensionality $n_v -1$
as its dual. The present graph-symmetry-based Hamiltonian, and
the resulting triangular lattice-like graphs in the low temperature phase, make
this interpretation unlikely.
The mean valency in the low temperature phase of our graph model is
approximately 6, corresponding to the triangular lattice graph; we thus
may consider a tesselation of a five-dimensional
Euclidean space by tetrahedra. The triangular lattice graph has smallest cycles of 3 vertices,
corresponding to traversing the smallest triangles in the graph.
However, a Euclidean tesselation using
non-obtuse
simplices
will have
cycles of its dual graph with no less than 4 vertices, i.e. due to the
acuteness (or more precisely non-obtuseness) of the
simplices,
every cycle consists of a (potentially non-planar) polygon of at least
4 sides. As an illustration of this, consider the dual graph to
a three-dimensional tesselation by tetrahedra with non-obtuse dihedral angles.
A section of a three-dimensional tiling by such
tetrahedra is shown in Fig.~}\hlref{\ref{FigCubeDual}(a),} \hll{and the
  corresponding dual graph is shown in Fig.~}\hlref{\ref{FigCubeDual}(b).}\hll{
Here we see that the smallest cycles of the dual graph are indeed 4,
corresponding to $\pi/2$ dihedral angles of the tiling tetrahedra.
However a significant fraction of the cycles have length 6. Moreover, the cycles
of length 4 appear as faces of 3D cubes in the dual lattice. All of this
structure is incompatible with a regular planar graph of
valency 4 as a potential
dual to the three-dimensional tesselation;
in particular, a graph having the topology of a square lattice is ruled
out.
}

\stt{Because there is no}
\hll{In the model, there are no}
constraints
on the global structure of the graph. \hll{As a consequence,} the low-temperature
phase can still retain
complicated topologies with small-world properties,
for which the corresponding manifold shows handles,
self-intersections, and local defects that deviate from the manifold,
in that a higher embedding dimension is necessary to represent them.
\hll{Defects on the low-temperature manifold induce scattering and lensing
  effects on the propagation of bosonic matter fields}\hlref{~\cite{Quach:2012tc},} \hll{and are an
interesting topic of future work for our model.}
\hll{As well, the presence of non-local links in the low temperature graph, and the
corresponding non-locality in the emergent manifold, is consistent
with the possible presence of disordered locality in loop quantum
gravity} \hlref{\cite{Markopoulou:2007ha},} \hll{
and might constitute a mechanism for its generation.
In the context of loop quantum gravity, macroscopic expectation values
of area or volume deviate from those on a flat metric by
$\mathcal{O}(\ell_p^2)$ or $\mathcal{O}(\ell_p^3)$ where $\ell_p$ is the
Planck length; nonlocal connections in the underlying metric modify
the local Hamiltonian coupling a matter field to loop quantum gravity,
but leave the above expectation values essentially unchanged,
indicating locality may be macroscopically smooth but microscopically
disordered.
}

As a general property of the graph model, the
high-temperature phase has an entropy that grows
  super-extensively with system size $N_V$.
This results in a transition temperature of zero
in the limit
$N_V \rightarrow \infty$, so that the infinite manifold is always
disordered at any finite temperature.
Aside from a finite universe or diverging coupling constraints as possible solutions,
we implemented long-range interactions
between vertex defects with repulsive Coulombic potential,
to energetically penalize the many graph configurations with defect
arrangements consistent with small-world topologies.
In analogy with low-dimensional condensed
matter systems, long-range potentials that couple defects induce
prohibitive energetic cost to configurations that would \hll{otherwise} destroy order
entropically, so that an ordered phase at low temperature \hll{is restored}.
Here, we found that such potentials result in a nearly
constant transition temperature
as the size of the graph $N_V$ increases. In addition, we found that attractive Coulombic potentials result in long-lived metastable states in the simulations.

Another interesting feature of the model is that the low lying energy
levels, including the ground state level,
have large configurational degeneracy. This residual entropy is due to
local defects that can ``absorb'' extra edges without energetic cost.
As well, the simulation dynamics
indicates that the energy barriers between different low-energy states
are not high. Thus at
temperatures below the phase transition, the degrees of freedom in the
system arising from this residual
entropy are not frozen. The small or zero-energy barriers between
degenerate states make the low-temperature graph system
similar to the spin ices observed in spinel structures and
pyrochlore lattices~\cite{Pauling:1935,AndersonPW56,BramwellST01}.

\hll{We have implemented here a sufficiently general Hamiltonian such that the
same dynamic model can give rise to space-times of different
dimensionality, i.e. spaces of different dimensions are solutions to
the same model. Exclusively from the graph theory point of view, there
is no} \hlref{{\it a priori}} \hll{reason to choose any particular dimensionality as a
phenomenological term in the Hamiltonian. The ``emergent
dimensionality'', then comes from initial conditions. Our motivation
for this was to choose the simplest Hamiltonian possible, that was
free of phenomenological parameters, so that the dimensionality of
space-time was not ``baked into'' the energy function that governed
dynamics.  That said, we acknowledge that this approach effectively
shifts the space-time dimension from extra phenomenological parameters
in the Hamiltonian that favor or disfavor particular subgraphs}
\hlref{\cite{Conrady:2010qz}}\hll{, to special initial conditions.}
\hll{Our Hamiltonian is local in that it is a sum over all the
vertices, and each term only depends on a small neighborhood (in our
case, the neighborhood subgraph) of each vertex. This contrasts with
other quantum graphity Hamiltonians, which have actions that depend on
the number of loops with long lengths}
\hlref{\cite{Konopka:2008hp}}.

It is intriguing to interpret the low-temperature configurations of this graph
model as an emergent spacetime --- a notion other researchers have explored for similar
graph
models~\cite{Ambjorn:2004qm,Ambjorn:2005qt,Konopka:2006hu,Konopka:2008hp,Konopka:2008ds,Hamma:2009xb,Caravelli:2011kq,Conrady:2010qz,Caravelli:2010xx}.
In this picture, general relativity is an effective
``hydrodynamic'' theory emerging from the collective dynamics of more
fundamental degrees of freedom. The graph model is appealing in that
both spacetime manifolds and locality emerge in the low-temperature
regime of a discrete structure.
The graph model introduced here gives rise to
real, positive distances, so the emergent manifold can only be
a \hll{Wick-rotated}, Euclidean version of spacetime.
\hll{
Monte Carlo ``time'' steps in the current Hamiltonian methodology are
distinct from the time evolution of the graph or manifold, and are
only a mechanism to sample equilibrium states.
In the present formulation, the Euclidean gravity theory undergoes a
phase transition to smooth metrics below a ``temperature'' parameter
$\beta$.
Exploiting the isomorphism between the quantum propagator and the
statistical mechanical partition function} \hlref{\cite{Kogut:1979wt},}
\hll{the quantity
$\mbox{e}^{-\beta \cal{H}}/\int [dg] \mbox{e}^{-\beta \cal{H}}$
is the equivalent to the Euclidean path integral measure that
determines Green functions $\left< g_1 \ldots g_n \right>$ for the
metric $g$ in a quantum gravity model with the corresponding action.
While we have seen a phase transition for the system with
Euclideanized action, the identification of the appropriate
thermal quantum states that are
periodic in real time, and so related to the parameter $\beta$, is not
clear at present. We see this problem of mapping back to the
space-time coordinates with Minkowskian signature as a
general challenge for quantum graphity models.
Another general issue is the absence of an underlying symmetry
principle to determine the action in quantum graphity models,
analogous to the role of general covariance in the action for quantum gravity. }

\stt{However, t}
The complex topologies of surfaces corresponding to low-temperature graphs, along with graph defects having zero energetic cost,
\hll{implies
that a graph model consisting solely of the current Hamiltonian} 
does not reduce to a
classical theory of Euclidean gravity in the macroscopic limit.
\hll{On the other hand, other discrete models of gravity are also known to have
scale-dependent spectral
dimension, indicating fractal, non-smooth geometries for the emergent
manifolds at least at intermediate length scales}
\hlref{\cite{Ambjorn:1995rg,Ambjorn:2005db,Giasemidis:2012qk}.}
The set of all possible low-energy
graphs in this model could potentially be identified with the phase space
of a \hll{Euclidean} gravity theory before imposing the equation of motion,
i.e., the space of all possible metrics modulo diffeomorphisms.
Because the low-temperature graphs of our model are nearly
triangulations, and random triangulations form the phase space of many
other discrete gravity
\hlref{models~\cite{DiFrancesco:1993nw,Johnston:1996rp,Bowick:1997za,Ambjorn:2005qt,Oriti:2006se},}
  it may be interesting to
investigate whether the graph model's action
may be extended to include terms in dynamical triangulation theory, which
do reduce to the gravitational action in the continuous limit.

\stt{The graph model in the present study possesses a
first order phase transition. It is possible that a combined model
such as those above
can have a second-order phase transition, or have two successive phase
transitions, in which one is the current first-order phase transition,
and the other new phase transition is second-order, corresponding to
the transition between complicated, fractal surface-like graphs and
extended surface-like graphs.}

\hll{The transition from disordered to ordered manifolds is first-order
  in the present graph model.
However,} the order of the transition, and its potential
relevance to
universality or independence of underlying lattice specifics, is a
non-issue for the investigation of ordered phases below the
transition, where correlation lengths are finite.
Power-law correlations calculated in causal dynamical triangulation are
between graphical elements analogous to
graviton fields, so that graviton coupling is power law as in the
classical limit. Space to time ratios of simplices have second order
transition in this model, while the transition involving gravitational coupling is
first order~\cite{Ambjorn:2011cg}. In any event, \hll{a graph model} at
a critical point would have wildly fluctuating connectivity
and resemble more a fractal mix of
ordered and disordered states, which is not consistent with an
emergent manifold.
The issue of the universality classes and corresponding exponents of a
transition is a separate one from the properties of an emergent
manifold as a low-temperature phase below a phase
transition. \hll{Retention of microscopic structure in the disorder to
  order transition is a prediction of the graph model,
  and may enable future experimental tests.}

Finally\stt{however}, it is intriguing to speculate on the utility
of a such a graph theoretical transition to describe a transition
involving non-local to local
causality, as might
occur in a pre-inflationary universe. Such models may address the
low-entropy initial condition problems that occur in inflationary
models \cite{Penrose:1988mg,Carroll:2005it}.

\section*{Acknowledgement}

We would like to thank Mark Van Raamsdonk and Moshe Rozali for helpful
discussions. S.C. is supported by the Faculty of Graduate Studies 4YF
Program at the University of British Columbia at
Vancouver. S.S.P. acknowledges support from the Canada Research Chairs
program. We also
acknowledge the Natural Sciences and Engineering Council for providing
funding to defray publication page fees.

\appendix
\section{The weighted histogram analysis method}

The weighted histogram analysis method (WHAM) is a method to combine
the samples from several Monte Carlo simulations taken under
conditions of different temperature and added potential. We employ WHAM to generate optimal estimates of energy distributions of the graph model.
In this model, the energy takes only integer values between $0$ and $M=1.5N_V$. Assume that $S$ simulations are
performed (in our cases, $S=4$ for $N_V=1000$, $S=10$ for $N_V=2000$), with inverse temperature $\beta_i$, and biasing potential $V_i(E)$, i.e., in the $i$-th simulation, the system is sampled with
energy distribution $\Omega(E)\exp(-\beta_i(E+V_i(E)))$, where
$\Omega(E)$ is the yet-unknown number of states with energy $E$. The
inverse temperature $\beta_i$'s are taken to be near the inverse
transition temperature. Because there
is a large free energy barrier between the low and high energy
phases near the transition temperature, a biasing
potential is used to obtain better sampling in the barrier region.
The form of the biasing potential is
taken to be parabolic:
\sen{
V_i(E)=\left\{\begin{array}{ll}
v_i\left[\frac{\left(E-\frac{E_i^l+E_i^h}{2}\right)^2}{\left(\frac{E_i^h-E_i^l}{2}\right)^2}-1\right],& E_i^l\le E\le E_i^h,\\
0,&E<E_i^l,\\
0,&E>E_i^h,
\end{array}\right.
}
where the parameters $v_i$, $E_i^l$ and $E_i^h$ are chosen by trial and error to make the energy distribution of each simulation as flat as possible.

After performing the simulations, let $n_i(E)$ be the number of counts of energy $E$ from the $i$-th
simulation, and $N_i$ the total number of samples from the $i$-th
simulation. From this information, the optimal estimate of the probability
$p^0(E)$ of energy level $E$ at inverse temperature $\beta^0$
without any biasing potential is given by
\se{
p^0(E)=\frac{\sum_{i=1}^Sn_i(E)}{\sum_{i=1}^SN_if_ic_i(E)},
\label{EqnWham1}
}
where $c_i(E)$ is the biasing factor $c_i(E)=\exp[-(\beta_i-\beta^0)E-\beta_iV(E)]$, and $f_i$ is a normalization constant satisfying
\se{
f_i^{-1}=\sum_{E=0}^M c_i(E)p^0(E).
\label{EqnWham2}
}
To solve these equations, we take an arbitrary set of initial values
for $f_i$ (namely $f_i^0=1$), and apply \eqref{EqnWham1} and \eqref{EqnWham2} iteratively to find the solution to these equations. After finding $p^0(E)$, it is then straightforward to calculate the average energy and heat capacity at inverse temperature $\beta^0$.

\section{\label{SecAcceptanceRatio}Acceptance ratio}

\begin{figure}
\centering
\includegraphics{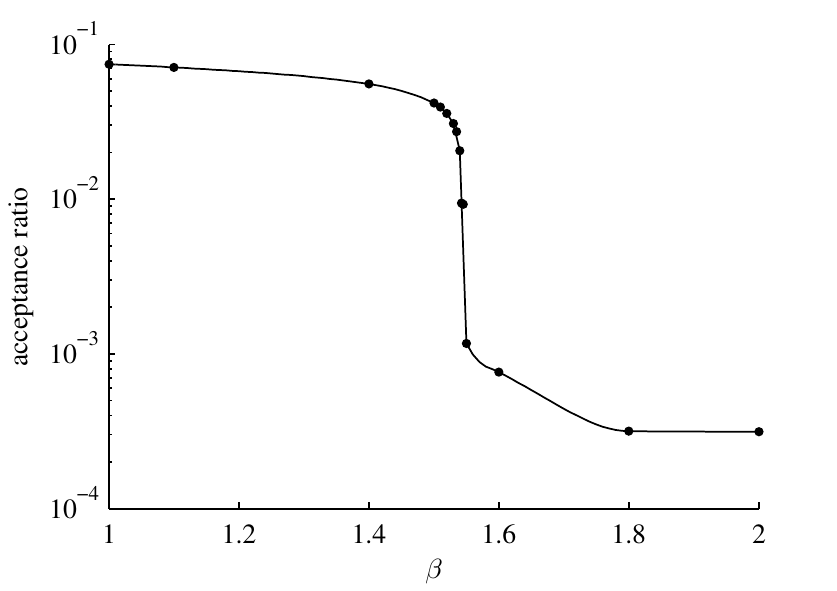}
\caption{\label{FigAccept} For the system of size $N_V=1000$, the acceptance ratio of Monte Carlo moves in the simulations is plotted as a function of inverse temperature $\beta$.}
\end{figure}

As a practical matter, we plot the acceptance ratio as a function of
$\beta$ for $N_V=1000$ in Fig.~\ref{FigAccept}. The low energy phase
occupied at large values of $\beta$ has a much lower acceptance ratio
than the high energy phase, both
because of the lower temperature and because the low energy graphs
have much more structural constraints, and thus have more rigidity with respect to
the local moves. However, because some the local defects cost little
or no energy, low energy graphs still have nonzero acceptance ratio,
and so are able to undergo dynamics during the simulations.


%

\end{document}